\documentclass[times,final]{elsarticle}

\usepackage{geometry}
\usepackage{framed,multirow}
\usepackage{url}
\usepackage{xcolor}
\definecolor{newcolor}{rgb}{.8,.349,.1}
\journal{Journal of Computational Physics}

\usepackage{graphicx}
\usepackage{subcaption}
\usepackage{amsmath,amsfonts,amssymb,amsthm, bm}
\usepackage{latexsym}
\usepackage{algorithm}
\usepackage{braket}
\usepackage{array}
\usepackage[noend]{algpseudocode}
\algrenewcommand\textproc{}

\begin{document}


\begin{frontmatter}

\title{Treecode-accelerated Green Iteration for Kohn-Sham Density Functional Theory}

\author[add1]{Nathan Vaughn}\corref{cor1}
\ead{njvaughn@umich.edu}
\cortext[cor1]{Corresponding author.}
\author[add2,add3]{Vikram Gavini}
\ead{vikramg@umich.edu}
\author[add1]{Robert Krasny}
\ead{krasny@umich.edu}

\address[add1]{Department of Mathematics, University of Michigan, Ann Arbor, Michigan 48109, USA}
\address[add2]{Department of Mechanical Engineering, University of Michigan, Ann Arbor, Michigan 48109, USA}
\address[add3]{Department of Materials Science and Engineering, University of Michigan, Ann Arbor, Michigan 48109, USA}
\date{\today}


\begin{abstract}
We present a real-space computational method
called treecode-accelerated Green Iteration (TAGI)
for all-electron Kohn-Sham Density Functional Theory.  
TAGI is based on a reformulation of the Kohn-Sham equations
in which the eigenvalue problem in differential form
is converted into a fixed-point problem in integral form
by convolution with the modified Helmholtz Green's function.
In each self-consistent field (SCF) iteration,
the fixed-points are computed by Green Iteration,
where the discrete convolution sums are efficiently evaluated 
by a GPU-accelerated barycentric Lagrange treecode.
Other techniques used in TAGI include
adaptive mesh refinement, Fej\'er quadrature,
singularity subtraction,
gradient-free eigenvalue update,
and
Anderson mixing to accelerate convergence of 
the SCF and Green Iterations.
Ground state energy computations of several atoms (Li, Be, O) 
and 
small molecules (H$_2$, CO, C$_6$H$_6$) 
demonstrate TAGI's ability to efficiently achieve chemical accuracy.
\end{abstract}


\end{frontmatter}

\section{Introduction}

Electronic structure calculations complement 
materials engineering experiments by predicting properties such as binding energy, 
inter-atomic forces, magnetization, and doping effects.  
Density Functional Theory (DFT)~\cite{hohenberg-kohn-1964}, 
which describes a system and its properties by its electron density, 
has been the workhorse of ground state electronic structure computations.  
For an $N_e$-electron system, 
the Kohn-Sham approach to DFT~\cite{kohn-sham-1965} reduces the 
$3N_e$-dimensional problem for the many-body wavefunction to a 
3-dimensional problem for the electron density.  
In particular,
the system of $N_e$ interacting electrons is replaced by a fictitious system of 
$N_e$ non-interacting electrons giving rise to the same electron density.
In principle, the Kohn-Sham formulation is exact for the 
ground state properties of materials systems, 
but it requires knowledge of the exchange-correlation functional, 
which is not known explicitly and is modeled in practice.
Approximating the exchange-correlation functional is an active area of research~\cite{Burke-2012,Jones-2015,Mardirossian-2017}, 
and better approximations enable Kohn-Sham DFT to more accurately predict 
ground state materials properties.  


{\bf Previous related work.}
There are many options for performing either 
all-electron or pseudopotential DFT calculations,
where, in the latter case, only the valence electrons are computed.
Often a basis set is used to represent the wavefunctions and electron density~\cite{lin-2019}.  
For periodic systems, the plane-wave basis is widely used for pseudopotential calculations~\cite{kresse-furthmuller-1996, gonze-2002, segall-2002, Giannozzi2009, Giannozzi2017}, 
and for all-electron calculations that require higher resolution to capture the rapidly oscillating wavefunctions, the augmented plane wave basis~\cite{loucks-slater-1967} and its variants are employed~\cite{Andersen1975,Wimmer1981,Weinert1982,Sjostedt2000,Madsen2001}.  
For non-periodic systems, 
Gaussian basis sets are widely used in quantum chemistry codes~\cite{g16, Nwchem2010} as they afford analytic evaluation of many integral and differential operators. 
A more recent option is the finite-element basis~\cite{Tsuchida1998,Pask1999,Lehtovaara2009,motamarri-nowak-leiter-knap-gavini-2013}, which efficiently treats periodic or non-periodic boundary conditions,
and
pseudopotential or all-electron systems using higher order finite-elements~\cite{motamarri-gavini-2014, kanungo-gavini-2017,motamarri-das-rudraraju-ghosh-davydov-gavini-2019,Das-2019}.

The previously described methods are based on solving the Kohn-Sham eigenvalue equation, 
a single-particle Schr\"{o}dinger-like differential equation. 
In this work we consider an alternative approach
in which the eigenvalue problem in differential form is converted into a 
fixed-point problem in integral form
by convolution with the modified Helmholtz Green's function.
While integral equation methods are extensively used for the wave equations arising in classical scattering~\cite{rokhlin-1990, jorgenson-mittra-1990, coifman-rokhlin-wandzura-1993, medgyesi-mitschang-1994, bleszynski-bleszynski-jaroszewicz-1996,volakis-sertel-2011,botha-2006} 
and quantum scattering~\cite{faddeyev-seckler-1963, johnson-secrest-1966, alt-grassberger-sandhas-1967,masel-merrill-miller-1975,adhikari-1986, hecht-2000}, these methods have received much less attention for eigenvalue problems 
corresponding to ground state calculations of the 
Schr\"{o}dinger or Kohn-Sham equations.
The integral equation approach was first applied by Kalos~\cite{kalos-1962} 
to solve the Schr\"{o}dinger equation for 3- and 4-electron systems
using Monte Carlo minimization.
Later, Zhao et al.~\cite{zhao-2007} used this approach to investigate various 
1-electron systems in 3D, 
where the convolution integrals were computed using the 
Multi-Level Fast Multipole Method.

In recent work, 
the integral equation approach was extended to the Hartree-Fock 
and Kohn-Sham equations, 
where the electron density was updated in self-consistent field (SCF) iterations,
and
the fixed-point problem for the wavefunctions and eigenvalues
in each SCF was solved by a process called Green Iteration.
Harrison et al.~\cite{harrison-2004}
implemented Green Iteration for the Kohn-Sham equations 
in a multiwavelet basis
that provides local refinement for each wavefunction,
and
this is now incorporated in the MADNESS code~\cite{harrison-2016}.
The convergence of Green Iteration for the many-body Schr\"{o}dinger equation 
was investigated by Mohlenkamp and Young~\cite{mohlenkamp-young-2008, mohlenkamp-2013}, who proved that the iteration converges for $N_e=1$ and $N_e=2$, 
provided the interaction potential belongs to the 
function space $L^2(\mathbb{R}^3) + L^{\infty}(\mathbb{R}^3)$ 
and 
the $L^{\infty}(\mathbb{R}^3)$ piece can be taken to be arbitrarily small. 
Khoromskij~\cite{khoromskij-2008} 
later extended this proof to Kohn-Sham DFT, where now the electron-electron interaction potential is replaced by the exchange-correlation potential which must satisfy the same function space requirements.  
Subsequently, 
Rakhuba and Oseledets~\cite{rakhuba-oseledets-2015,rakhuba-oseledets-2016}
applied Green Iteration to the Hartree-Fock and Kohn-Sham equations 
in a Tucker tensor basis that uses low rank approximations of the wavefunctions.

{\bf Present work.}
We present a new integral equation based method called 
Treecode-Accelerated Green Iteration (TAGI)
for all-electron Kohn-Sham DFT calculations.
The key features of TAGI that enable accurate and efficient 
calculations are 
(1) adaptive mesh refinement,
(2) high order quadrature,
(3) singularity subtraction for convolution integrals,
(4) gradient-free eigenvalue update,
(5) Anderson mixing for SCF and Green Iteration, and
(6) treecode computation of discrete convolution sums.

TAGI is a real-space method in which the fields are represented 
directly at quadrature points. 
TAGI uses adaptive mesh refinement to efficiently represent the fields, 
which vary rapidly near the nuclei but decay smoothly in the far-field.
The adaptive refinement scheme results in a set of cuboid cells, 
which are discretized with Chebyshev points of the first kind, 
and 
all integrals are evaluated with the Fej\'er (``classical'' Clenshaw-Curtis) quadrature rule~\cite{fejer-1933,trefethen-2008b}.
The convolution integrals have singular kernels (Coulomb and Yukawa), 
which impede the accuracy of the quadrature rule,
and
we employ singularity subtraction to reduce the error in the quadrature sums.
A standard singularity subtraction scheme is used for the 
Yukawa kernel~\cite{rabinowitz-1961, anselone-1981}  
and 
we developed a modified version for the Coulomb kernel.
To further improve accuracy, 
we use a gradient-free eigenvalue update~\cite{harrison-2004} 
within Green Iteration to eliminate the error arising from 
numerical differentiation in the standard gradient eigenvalue update. 
We analyze the convergence rate of Green Iteration 
and 
use a fixed-point acceleration technique to alleviate slow convergence.
Finally, the discrete convolution sums are efficiently evaluated
using a Barycentric Lagrange Treecode~\cite{wang-krasny-tlupova-2019} (BLTC),
which reduces the computational complexity from $O(N^2)$ to $O(N\log N)$ 
while introducing a small and controllable approximation error.  
Furthermore, the BLTC is accelerated on GPUs with OpenACC~\cite{vaughn-wilson-krasny-2020} and across multiple GPUs on a  single node with OpenMP.
We demonstrate the impact on accuracy and efficiency of each of the previously described features on the carbon monoxide molecule,
and
then perform ground state energy calculations for several atoms and molecules, 
demonstrating TAGI's ability to achieve chemical accuracy of 1 mHa/atom.

The paper is organized as follows.
Section~\ref{section:Kohn-Sham formulation} presents Kohn-Sham DFT, and the standard Self-Consistent Field iteration for computing the ground state density and wavefunctions. 
Section~\ref{section:integral-form} 
presents the integral equation formulation we employ
and Green Iteration for the resulting fixed-point problem.  
Section~\ref{section:spatial-discretization} describes the numerical techniques developed in this work to enhance the accuracy of the integral formulation, and demonstrates these ideas on the carbon monoxide molecule.
Section~\ref{section:wavefunction-mixing} investigates the convergence rate of Green Iteration and demonstrates the fixed-point acceleration technique used in TAGI.  
Section~\ref{section:treecode} describes the treecode algorithm 
for computing fast approximations of the convolution integrals
and demonstrates the efficiency of the GPU-accelerated implementation used in this work.
Section~\ref{section:Total-Energy-Accuracy} 
applies TAGI to several atoms and small molecules, 
achieving chemical accuracy of 1~mHa/atom with respect to reference values.  
Section~\ref{section:Conclusion} provides a summary of our findings, and discusses a path forward for this 
approach to further improve performance and scale to larger systems.


\section{Kohn-Sham Density Functional Theory} \label{section:Kohn-Sham formulation}

The input to Kohn-Sham DFT consists of 
the positions and atomic numbers of the atoms in the system,
and the output consists of the ground-state electron density
along 
with the Kohn-Sham single-electron wavefunctions, 
from which the desired observables 
(including ground-state energy and ionic forces) can be computed.
The Kohn-Sham equations are 
\begin{equation}
\mathcal{H}[\rho]\psi_i({\bf r}) = 
\varepsilon_i\psi_i({\bf r}), \quad i=1, 2, \ldots, \quad
\mathcal{H}[\rho] = -\frac{1}{2}\nabla^2 + V_{eff}[\rho]\,,
\label{eqn:Kohn-Sham-differential}
\end{equation}
where 
$\mathcal{H}[\rho]$ is the Kohn-Sham Hamiltonian,
$\rho = \rho({\bf r})$ is the electron density,
$\varepsilon_i$ are the Kohn-Sham eigenvalues,
and
$\psi_i({\bf r})$ are the Kohn-Sham eigenfunctions, also referred to as the Kohn-Sham wavefunctions. Here, we restrict ourselves to a spin-independent formulation on non-periodic systems, 
but the general ideas presented in this work can be extended to a 
spin-dependent formulation and periodic geometries in a straightforward manner. 
The effective Kohn-Sham potential has the form,
\begin{equation}
\label{eqn:effective-potential}
V_{eff}[\rho]({\bf r}) = V_{H}[\rho]({\bf r}) + V_{ext}({\bf r}) + V_{xc}[\rho]({\bf r}),    
\end{equation}
where the first two terms are
the Hartree potential due to the electron density
and
the external potential due to the $N_A$ atomic nuclei
located at ${\bf R}_j$ with charges $Z_j$, respectively,
\begin{equation}
V_{H}[\rho]({\bf r)} = \int\frac{\rho(\mathbf{r}^\prime)}{|\mathbf{r}-\mathbf{r}^\prime|}d\mathbf{r}^\prime, \quad
V_{ext}(\mathbf{r}) = \sum_{j=1}^{N_A} \frac{-Z_j}{|\mathbf{r}-\mathbf{R}_j|},
\label{eqn:classical-electrostatics}
\end{equation}
and
the third term is the exchange-correlation potential $V_{xc}[\rho] = \partial E_{xc}[\rho]/\partial\rho$
depending on the exchange-correlation energy $E_{xc}[\rho]$. 
The electron density depends on the eigenvalues and wavefunctions,
\begin{equation}
\rho(\mathbf{r}) = 2\sum_{i=1}^{N_w}f(\varepsilon_i,\mu_F)|\psi_i(\mathbf{r})|^2, \quad
f(\varepsilon,\mu_F) = \frac{1}{e^{(\varepsilon-\mu_F)/{k_BT}}+1},
\label{eqn:density-construction}
\end{equation}
where 
$f(\varepsilon,\mu_F)$ is the fractional occupation computed by Fermi-Dirac
statistics~\cite{kresse-furthmuller-1996, goedecker-1999}, 
with Fermi energy $\mu_F$, 
Boltzmann constant $k_B$, and temperature $T$.  
The Fermi energy $\mu_F$ is determined
from the constraint on the total number of electrons $N_e$,
\begin{equation}
2\sum_{i=1}^{N_w}f(\varepsilon_i,\mu_F) = N_e\,. 
\label{eqn:Fermi}
\end{equation}
The sums in Eq.~\eqref{eqn:density-construction} and Eq.~\eqref{eqn:Fermi} 
run over the $N_w$ lowest energy wavefunctions,
where $N_w$ is chosen so that the fractional occupation of any
higher energy wavefunction is negligible.

The preceding equations constitute a non-linear eigenvalue problem
and
the standard solution method uses
the Self-Consistent Field iteration (SCF) outlined in Algorithm~\ref{alg:SCF}.  
The iteration takes the atomic positions and an initial guess for the electron density as input.
The output is the converged electron density and wavefunctions, from which observables are computed.
The iteration starts in line 1.
In line 2, at the $n$th step of the iteration,
the effective potential $V_{eff}[\rho_{in}^{(n)}]$ 
is constructed from the input electron density of the current iterate by Eq.~\eqref{eqn:effective-potential}.
In line 3, the eigenvalue problem in Eq.~\eqref{eqn:Kohn-Sham-differential}
is solved for the eigenpairs ($\varepsilon_i, \psi_i$).
In line 4, the Fermi energy $\mu_F$
and
fractional occupations $f(\varepsilon_i,\mu_F)$ are computed.
In line 5,
these quantities are used to compute a new output density $\rho_{out}^{(n)}$ 
by Eq.~\eqref{eqn:density-construction}. 
In line 6, the scheme checks whether the density has converged to a desired tolerance;
if so, then the iteration stops and returns the latest density;
otherwise a new input density $\rho_{in}^{(n+1)}$ is constructed by 
Anderson mixing~\cite{anderson-1965}
and
the iteration continues.
The present work follows this approach,
but focuses on the solution of the eigenvalue problem (line 3),
which is the most computationally intensive step in the SCF iteration,
using treecode-accelerated Green Iteration (TAGI) described below.

\begin{algorithm}
\caption{Self-Consistent Field Iteration (SCF)}\label{alg:SCF}
{\bf input: }atomic positions and initial guess for electron density $\rho_{in}^{(0)}$\\
{\bf  output:} electron density $\rho_{out}^{(n)}$ and Kohn Sham wavefunctions $\psi_i^{(n)}, i=1,\ldots, N_w$
\begin{algorithmic}[1]
\State for $n=0,1,2,\ldots$
\State \quad 
given $\rho_{in}^{(n)}$, construct effective potential $V_{eff}[\rho_{in}^{(n)}]$ 
by Eq.~\eqref{eqn:effective-potential}
\State \quad 
using $V_{eff}[\rho_{in}^{(n)}]$,
solve eigenvalue problem
$\mathcal{H}[\rho_{in}^{(n)}]\psi_i^{(n)} = \varepsilon_i^{(n)}\psi_i^{(n)}, i=1,\ldots, N_w$
\State \quad 
using $\varepsilon_i^{(n)}$, compute Fermi energy $\mu_F$
and
fractional occupations $f(\varepsilon_i^{(n)},\mu_F)$ by Eq.~\eqref{eqn:Fermi}
\State \quad 
using $f(\varepsilon_i^{(n)},\mu_F), \psi_i^{(n)}$,
construct new density $\rho_{out}^{(n)}$ by Eq.~\eqref{eqn:density-construction}
\State \quad if {$||\rho_{out}^{(n)} - \rho_{in}^{(n)}||_2 < tol_{scf}$ }, return $\rho_{out}^{(n)}$
\State \quad else construct new density $\rho_{in}^{(n+1)}$ by Anderson mixing
and return to step 2
\end{algorithmic}

\end{algorithm}

Having obtained the converged $\varepsilon_i, \psi_i({\bf r}), \rho({\bf r})$,
the ground-state energy of the system is
\begin{equation}
E = 
E_{kin} + E_{xc} + E_{H} + E_{ext} + E_{ZZ}.
\label{eqn:total-energy-gradient}
\end{equation}
In this expression,
the first two terms are
the kinetic energy and exchange-correlation energy, respectively,
\begin{equation}
E_{kin} = 
\sum_{i=1}^{N_w}\displaystyle\int \psi_i(\mathbf{r})\left(-\frac{1}{2}\nabla^2\right) \psi_i(\mathbf{r})d\mathbf{r}, \quad
E_{xc}[\rho] = \displaystyle\int \varepsilon_{xc}[\rho](\mathbf{r})\rho(\mathbf{r})d\mathbf{r},
\label{eqn:kinetic-energy}
\end{equation}
where $\varepsilon_{xc}[\rho]({\bf r})$
is the exchange-correlation energy per electron for the chosen DFT functional,
and
the remaining three terms are the
Hartree energy, external electrostatic energy, and nuclear repulsion energy, respectively,
\begin{equation}
E_{H}[\rho] = \frac{1}{2}\displaystyle\int V_H[\rho](\mathbf{r})\rho(\mathbf{r})d\mathbf{r}, \quad
E_{ext}[\rho] = \displaystyle\int V_{ext}(\mathbf{r})\rho(\mathbf{r})d\mathbf{r}, \quad
E_{ZZ} = \frac{1}{2}\sum_{i,j\neq i} \frac{Z_i Z_j}{|{\bf R}_i - {\bf R}_j|}.
\label{eqn:electrostatics}
\end{equation}
This work employs the Local Density Approximation 
(LDA)~\cite{ceperley-alder-1980,perdew-zunger-1981}
for $V_{xc}[\rho], \varepsilon_{xc}[\rho]$ 
which are computed using the \textit{Libxc} package~\cite{marques-2012,lehtola-2018}.

\section{Solution of Eigenvalue Problem by Green Iteration}

\label{section:integral-form}
Several methods are available for solving the eigenvalue problem in each SCF iteration 
(step~4 in Algorithm~\ref{alg:SCF}). 
Among real-space methods, finite-difference~\cite{Chelikowsky1994,bernholc-2008,saad-2010} and finite-element~\cite{motamarri-nowak-leiter-knap-gavini-2013,motamarri-das-rudraraju-ghosh-davydov-gavini-2019} methods represent the differential operator as a sparse matrix 
and 
use iterative techniques to compute the eigenpairs $(\varepsilon_i,\psi_i)$. 
By contrast,
in this work
the differential equation is converted into an integral equation 
by convolution with the modified Helmholtz Green's function~\cite{kalos-1962},
and
then an iterative technique called Green Iteration is applied 
to obtain the eigenpairs~\cite{harrison-2004,mohlenkamp-young-2008,khoromskij-2008,rakhuba-oseledets-2015}. 
We describe these steps below.

Following Kalos~\cite{kalos-1962}, 
the Kohn-Sham equations in Eq.~\eqref{eqn:Kohn-Sham-differential} 
are rewritten in the form
\begin{equation}
\left(\frac{1}{2}\nabla^2 + \varepsilon_i\right)\psi_i = V_{eff}[\rho]\psi_i,
\label{eqn:Helmholtz_Form}
\end{equation}
where $\rho$ is the electron density for a given SCF iteration.
Since the bound state eigenvalues of the Kohn-Sham Hamiltonian are negative,
$\varepsilon_i < 0$, 
Eq.~\eqref{eqn:Helmholtz_Form} is a modified Helmholtz equation
with Green's function,
\begin{equation}
G_{\varepsilon_i}(\mathbf{r},\mathbf{r}^\prime) = -\frac{e^{-\sqrt{-2\varepsilon_i}|\mathbf{r}-\mathbf{r}^\prime|}}{2\pi|\mathbf{r}-\mathbf{r}^\prime|},
\label{eqn:Hemlholtz_Greens_Function}
\end{equation}
where free-space boundary conditions are assumed. 
Then convolution with Eq.~\eqref{eqn:Helmholtz_Form} yields 
the integral form of the Kohn-Sham eigenvalue problem,
\begin{equation}
\psi_i(\mathbf{r}) = 
\mathcal{G}(\varepsilon_i)\psi_i({\bf r}),
\quad i = 1,\ldots, N_w,
\label{eqn:Integral_Form}
\end{equation}
where 
\begin{equation}
\mathcal{G}(\varepsilon)\psi({\bf r}) =
\int G_{\varepsilon}({\bf r},{\bf r}^\prime)
V_{eff}[\rho](\mathbf{r}^\prime)\psi(\mathbf{r}^\prime)d\mathbf{r}^\prime,
\label{eqn:integral_operator}
\end{equation}
defines a 1-parameter family of linear integral operators.
 
Note that Eq.\eqref{eqn:Integral_Form} can be viewed as a fixed-point problem
and
this motivates the solution method called Green Iteration
described in Algorithm~\ref{alg:GreensIteration}.
The scheme takes as input the effective potential $V_{eff}[\rho]$ for the current SCF
and
an initial guess for the eigenpairs $(\varepsilon_i^{(0)},\psi_i^{(0)})$,
and
provides the converged eigenpairs $(\varepsilon_i,\psi_i)$ as output.
Line 1 is the outer loop over wavefunctions
and 
line 2 is the iteration for a given wavefunction.
Line 3 applies the integral operator $\mathcal{G}(\varepsilon_i^{(n)})$
to the current wavefunction $\psi_i^{(n)}$.
Line 4 updates the eigenvalue;
several methods are available and we compare some of them below.  
Line 5 is the deflation step
that orthogonalizes the new wavefunction $\psi_i^{(n+1)}$ 
against the previously converged wavefunctions,
and line 6 normalizes it.
Line 7 checks for convergence;
if the tolerance is satisfied,
then the eigenpair is stored
and
the process returns to line 1;
otherwise the iteration in line 2 continues.

\begin{algorithm}\caption{Green Iteration}
\label{alg:GreensIteration}
{\bf  input}: effective potential $V_{eff}[\rho]$ for current SCF \\
{\bf input}: initial guess for eigenpairs $(\varepsilon_i^{(0)},\psi_i^{(0)}), i=1,\ldots,N_w$ \\
{\bf  output}: eigenpairs $(\varepsilon_i,\psi_i), i=1,\ldots,N_w$\begin{algorithmic}[1]
\State for $i=1, 2, \ldots, N_w$
\State \quad for $n = 0, 1, 2, \ldots$
\State \quad \quad compute $\psi_i^{(n+1)} = \mathcal{G}(\varepsilon_i^{(n)})\psi_i^{(n)}$
\State \quad \quad update eigenvalue $\varepsilon_i^{(n+1)}$
\State \quad \quad orthogonalize $\psi_i^{(n+1)}$ against previously converged
wavefunctions $\psi_j, j<i$
\State \quad \quad normalize $\psi_i^{(n+1)}$
\State \quad \quad if $||\psi_i^{(n+1)} - \psi_i^{(n)}||_2 < tol_{gi}$
set $(\varepsilon_i, \psi_i) = (\varepsilon_i^{(n)}, \psi_i^{(n)})$ 
and return to line 1
\State \quad \quad else return to line 2 and continue iteration
\end{algorithmic}
\end{algorithm}


\section{Spatial Discretization Techniques} \label{section:spatial-discretization}

This section focuses on the spatial discretization techniques used in TAGI.  
These include
the initialization scheme for the electron density and wavefunctions,
the quadrature and adaptive mesh refinement techniques,
the singularity subtraction schemes used to evaluate the convolution integrals,
and
the gradient-free approach used to update the eigenvalues.
The section concludes by
demonstrating the effect of these techniques 
using the carbon monoxide molecule as an example.
Note that Hartree atomic units are used and a table containing the physical and numerical parameters is provided in~\ref{appendix:parameters}.


\subsection{Initial Electron Density and Eigenpairs}
\label{section:initialization}
The SCF iteration uses an initial guess for the electron density of the form,
\begin{equation}
\rho^{(0)}({\bf r}) = \sum_{j=1}^{N_a} \rho_j(|{\bf r} - {\bf R}_j|), 
\label{eqn:initial_density}
\end{equation}
where $\rho_j(|{\bf r} - {\bf R}_j|)$ is a radial 1-atom electron density associated with the $j$th atom.
These 1-atom densities are precomputed by solving a radial version of the Kohn-Sham problem 
for each atomic species.
In addition, Green Iteration requires an initial guess for the eigenpairs, 
$(\varepsilon_i^{(0)}, \psi_i^{(0)}({\bf r})),i=1,\ldots,N_w$.
The number of wavefunctions $N_w$ is determined as follows.
Since each wavefunction is occupied up to two electrons, 
there is a lower bound, $N_w \ge N_e/2$,
however there is no sharp upper bound.
In practice $N_w$ should be chosen large enough to accommodate all states with
significant fractional occupation $f(\varepsilon_i,\mu_F)$.
To this end, $N_w$ is initialized to be larger than $N_e/2$, 
and
upon obtaining the eigenpairs, 
if the fractional occupation of the highest state is negligibly small, 
then $N_w$ is considered large enough;  
otherwise, $N_w$ is increased and the process is repeated until the
check is satisfied.
The initial guess for the eigenpairs depends on whether or not this is the first step in the SCF iteration. 
In the first step,
the wavefunctions are initialized using 1-atom wavefunctions obtained in the radial solve for 
the initial electron density, multiplied by appropriate spherical harmonics,
and
the initial eigenvalues are computed by the Rayleigh quotient,
$\varepsilon_i^{(0)} = \langle \psi_i^{(0)}, \mathcal{H}\psi_i^{(0)} \rangle$. 
In subsequent steps of the SCF iteration, 
the eigenpairs of the previous step are taken as the initial guess.


\subsection{Spatial Discretization and Quadrature Schemes} \label{section:quadrature}

\begin{figure}[htb]
\centering
\includegraphics[width=0.325\textwidth]{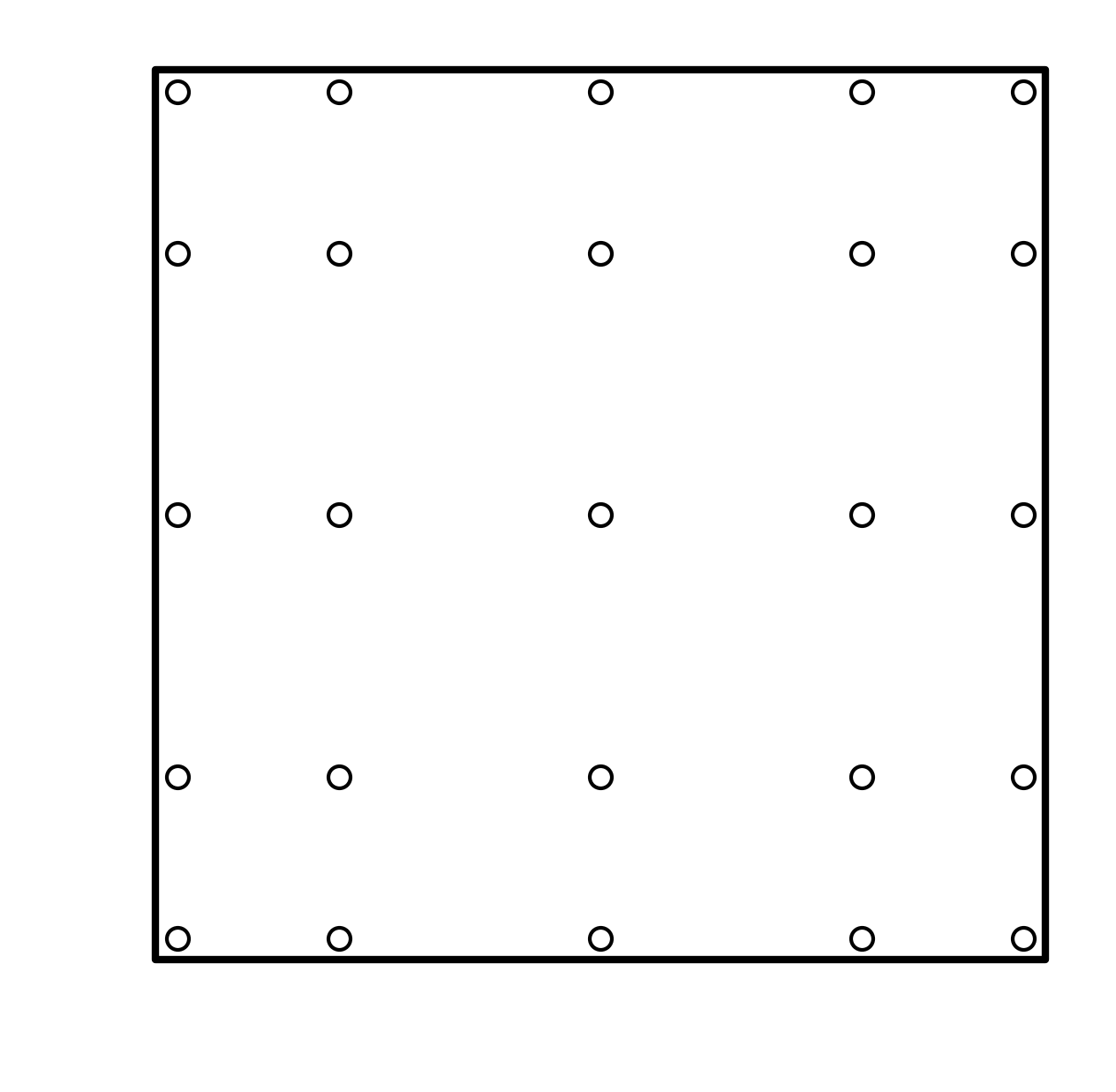}
\caption{
\textmd{A tensor product grid of Chebyshev points of the first kind 
in Eq.~\eqref{eqn:Chebyshev_points} with $p=4$ in a 2D cell.}
}
\label{fig:quadrature-points}
\end{figure}

The energy integrals and convolution integrals will be evaluated on a set of cuboid cells
representing a bounded computational domain.
Using the Hartree energy as an example,
\begin{equation}
E_H = \frac{1}{2}\int V_H({\bf r})\rho({\bf r}) d{\bf r} \approx
\frac{1}{2}\sum_{i=1}^{N_c} \int_{C_i} V_H({\bf r})\rho({\bf r}) d{\bf r} \approx
\frac{1}{2}\sum_{i=1}^{N_c} \sum_{j=1}^{(p+1)^3} V_H({\bf r}_{ij})\rho({\bf r}_{ij}) w_{ij},
\label{eqn:quadrature-example}
\end{equation}
where 
$N_c$ is the number of cells,
$(p+1)^3$ is the number of quadrature points in each cell,
indices $i,j$ refer to quadrature point $j$ in cell $i$,
and
$w_{ij}$ are the quadrature weights.
The total number of mesh points is denoted by $N_m = (p+1)^3 N_c$.
The quadrature scheme uses Chebyshev points of the first kind;
on the interval $[-1,1]$ these are given by
\begin{equation}
x_i = \cos\theta_i, \quad \theta_i = \frac{(2i+1)\pi}{2p+2}, \quad i = 0:p.
\label{eqn:Chebyshev_points}
\end{equation}
A tensor product grid of $(p+1)^3$ Chebyshev points is adapted to each cell;
Fig.~\ref{fig:quadrature-points} shows a 2D schematic.
Note that the Chebyshev points 
lie entirely inside the cell and never coincide with a vertex;
as explained below this is important because
the cells are chosen so that the atoms are located at cell vertices,
thereby avoiding the singularity of the nuclear potential.
Within each cell, the integrals are evaluated using the Fej\'{e}r 
(or ``classical'' Clenshaw-Curtis) quadrature rule~\cite{fejer-1933,trefethen-2008b}
with quadrature weights $w_{ij}$.
The $p+1$ point Fej\'{e}r quadrature rule integrates $p$th-degree polynomials exactly, 
so we refer to this as a $p$th-order quadrature rule.

\begin{figure}[htb]
\centering
\begin{subfigure}[b]{0.225\textwidth}
\centering
\includegraphics[width=\textwidth]{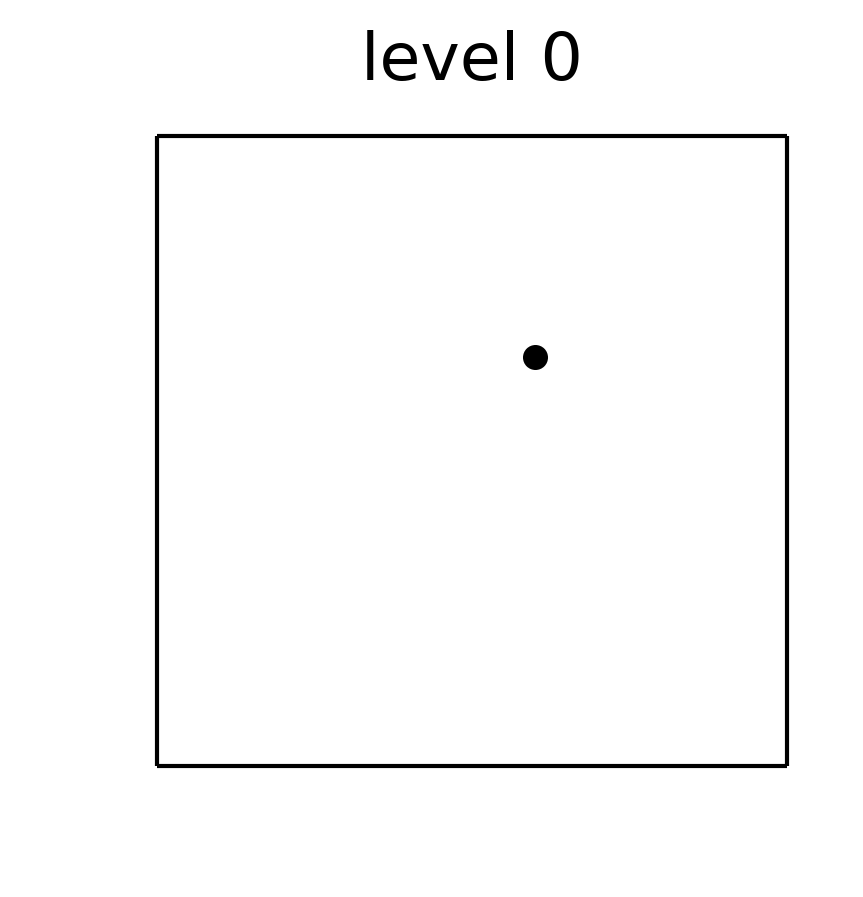}
\end{subfigure}
\begin{subfigure}[b]{0.225\textwidth}  
\centering 
\includegraphics[width=\textwidth]{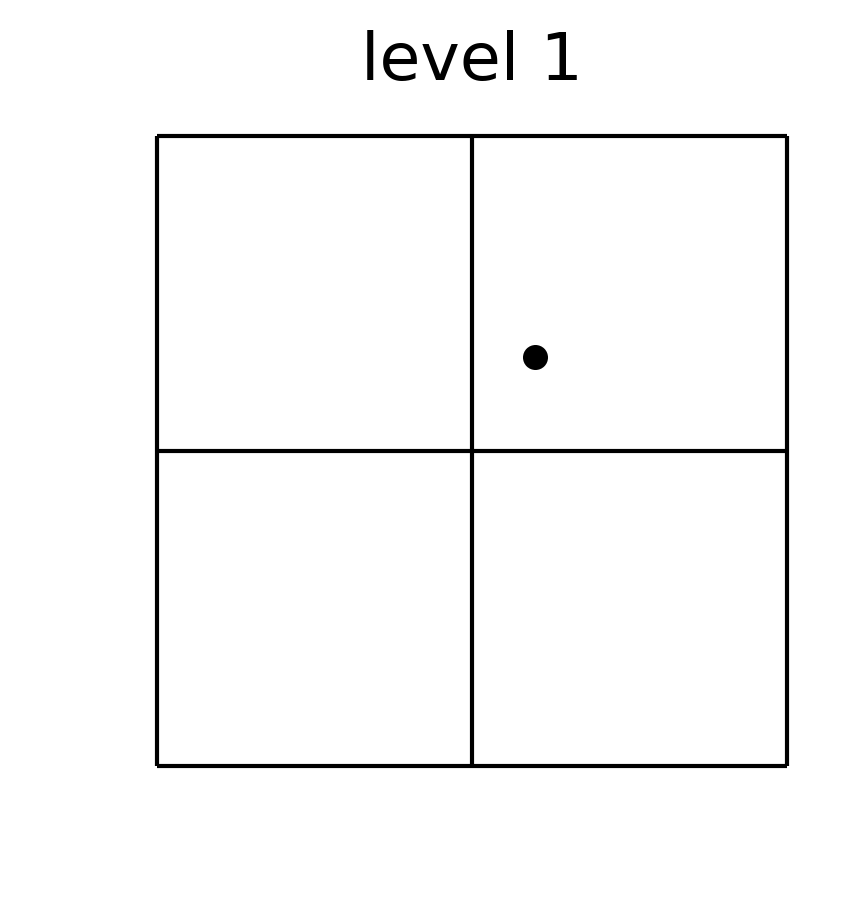}
\end{subfigure}
\begin{subfigure}[b]{0.225\textwidth}   
\centering 
\includegraphics[width=\textwidth]{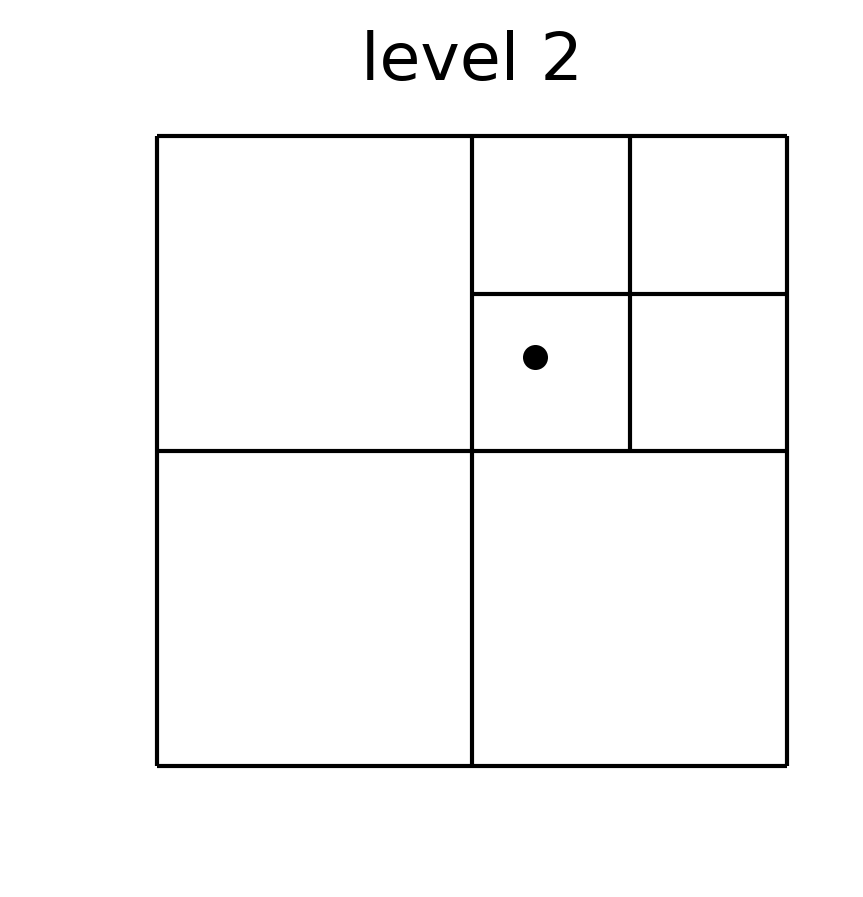}
\end{subfigure}
\begin{subfigure}[b]{0.225\textwidth}   
\centering 
\includegraphics[width=\textwidth]{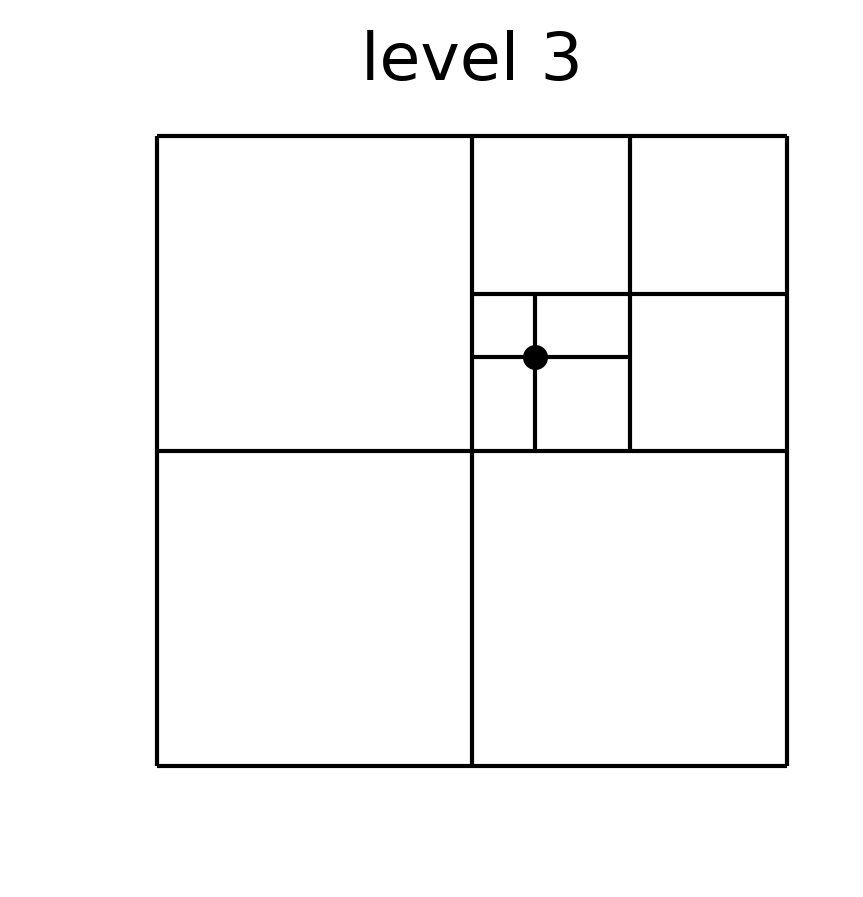}
\end{subfigure}
\caption{
\textmd{Illustration of the adaptive refinement scheme for a 1-atom system with the
atom located at ($\bullet$).
Four levels of refinement are shown where
the final refinement level puts the atom at a cell vertex.}
} 
\label{fig:spatial-discretization}
\end{figure}

The cells are defined using an adaptive refinement scheme illustrated in Fig.~\ref{fig:spatial-discretization}
for a 1-atom example.
The goal of the scheme is to produce cells that resolve the regions with 
significant electron density and wavefunction variation,
primarily near the atoms.
Level 0 is a large cube surrounding the atoms in the system,
with dimensions chosen to ensure that the electron density and wavefunctions 
are sufficiently small at the boundary.
The cube is refined by bisecting it in the three coordinate directions,
resulting in eight child cells.
Several levels of uniform refinement are performed,
and
subsequent refinement is done adaptively in the following manner.
Given a cell $C$,
we temporarily create the child cells $C_i, i=1:8$, 
and check the following criterion,
\begin{equation}
\left|\,\int_C t({\bf r})d\mathbf{r} ~-\, \sum_{i=1}^8 \int_{C_i} 
t({\bf r})d\mathbf{r}\,\right| < tol_m,
\label{eqn:refinement_criterion}
\end{equation}
where $t({\bf r})$ is a test function specified below
and
$tol_m$ is a user-specified tolerance.
The integrals in Eq.~\eqref{eqn:refinement_criterion} are evaluated using the Fej\'er rule.
If Eq.~\eqref{eqn:refinement_criterion} is satisfied,
then refinement is not needed and the child cells are discarded; 
otherwise the child cells are retained and the process continues.
Figure~\ref{fig:spatial-discretization} shows the schematic of a possible outcome 
where the initial cell is refined at level 1,
but only the child cell containing the atom is refined at level 2.
Once the tolerance is satisfied for every cell,
a final refinement step occurs;
those cells containing an atom are subdivided so that the atoms lie at cell vertices; 
this ensures that the Chebyshev grid points never coincide with an atom position
and
hence the fields (effective potential, wavefunctions, electron density) 
are smooth on the interior of the cells.
If the refinement scheme creates any cells with large aspect ratio, 
these cells are refined along their longest dimension.  

Several options for the refinement test function were considered
and
we decided to use 
\begin{equation}
t(\mathbf{r})=\sqrt{\rho^{(0)}(\mathbf{r})}V_{ext}(\mathbf{r}),
\label{eqn:test_function}
\end{equation}
where $\rho^{(0)}({\bf r})$ is the initial electron density in Eq.~\eqref{eqn:initial_density}
and
$V_{ext}(\mathbf{r})$ is the external potential in Eq.~\eqref{eqn:classical-electrostatics}.
This choice is motivated by several considerations.
First,
it resembles the function $\psi(\mathbf{r})V_{eff}(\mathbf{r})$
appearing in the integral form of the Kohn-Sham equations~\eqref{eqn:integral_operator};
this is because near a nucleus, 
$\sqrt{\rho^{(0)}(\mathbf{r})}$ has the characteristics of an $s$-orbital atomic wavefunction,
capturing the cusp and decay rate,
and
although $V_{eff}(\mathbf{r})$ is not known,
$V_{ext}(\mathbf{r})$ is known
and
contains the Coulomb singularities that must be resolved.
Second,
this test function is accessible at the start of the computation
and
can be evaluated at arbitrary grid points as needed in the refinement scheme.

\begin{figure}[htb]
\centering
\begin{subfigure}[b]{0.475\textwidth}
\centering
(a) $tol_m =$ 1e$-4$
\includegraphics[width=\textwidth]{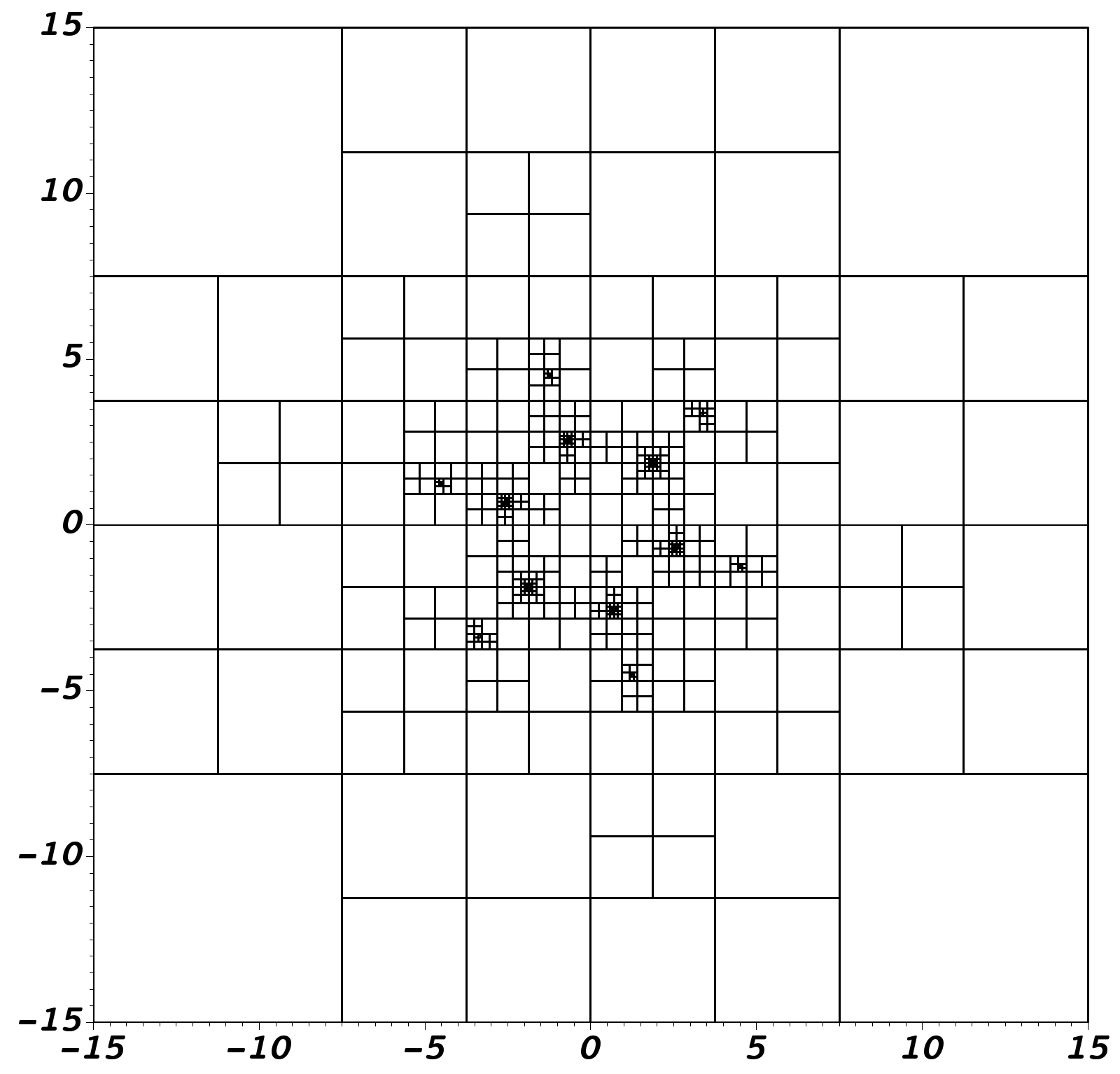}
\end{subfigure}
\begin{subfigure}[b]{0.475\textwidth}    
\centering 
(b) $tol_m =$ 3e$-6$
\includegraphics[width=\textwidth]{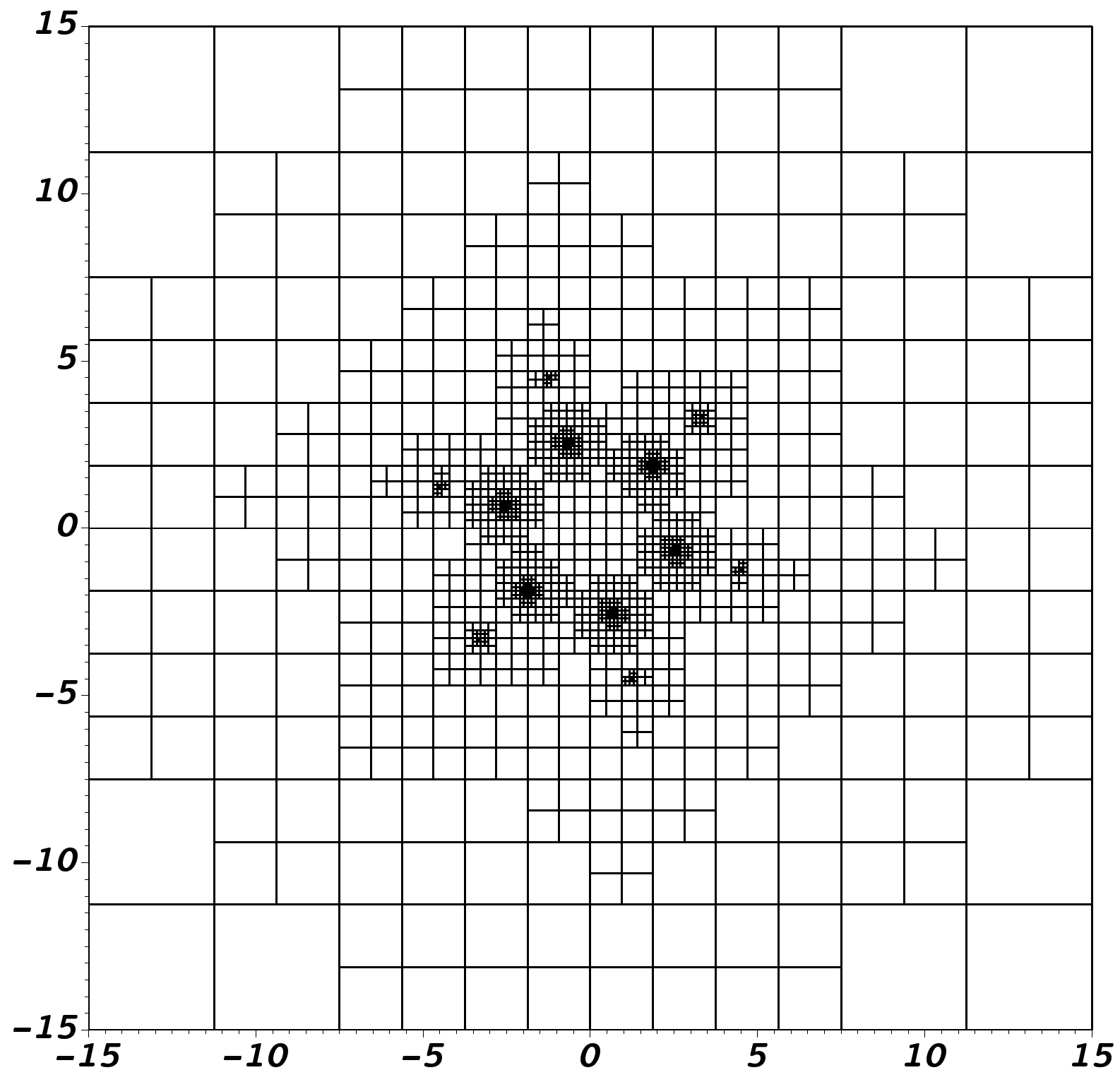}
\end{subfigure}
\caption{
\textmd{
Example of the mesh refinement scheme for the benzene molecule (C$_6$H$_6$).
2D slices of the mesh are shown in the plane of the molecule generated with
4th order quadrature in Eq.~\eqref{eqn:test_function} and
(a)~$tol_m=$1e$-4$,
(b)~$tol_m=$3e$-6$.} 
}
\label{fig:mesh-refinement-sequence}
\end{figure}

Figure~\ref{fig:mesh-refinement-sequence} shows an example of 
coarse and fine meshes for the benzene molecule (${\rm C}_6{\rm H}_6$)
obtained using the refinement scheme described above with 4th order quadrature.
The molecule lies in the $z=0$ plane
and
a truncated portion of the mesh in that plane is shown.
The coarse mesh is generated with $tol_m=1{\rm e}{-}4$ 
and 
the fine mesh with $tol_m=3{\rm e}{-}6$.
The resulting cell density is highest near the twelve nuclei,
and 
the carbon atoms are more highly refined than the hydrogen atoms, 
as expected since the test function $\sqrt{\rho^{(0)}(\mathbf{r})}V_{ext}(\mathbf{r})$ 
grows faster at heavier nuclei.
Compared to a variety of other refinement schemes we considered, 
this approach gave the best combination of accuracy and efficiency.
Further below we will demonstrate convergence with respect to 
both the order of the quadrature rule $p$
and the mesh tolerance parameter $tol_m$.


\subsection{Singularity Subtraction}
\label{section:singularity-subtraction}

Achieving the necessary accuracy for DFT calculations requires careful treatment of the 
singular integrals arising in Green Iteration,
\begin{equation}
    \psi^{(n+1)}(\mathbf{r}) = -\int  V_{eff}(\mathbf{r}^\prime)\psi^{(n)}(\mathbf{r}^\prime)\frac{e^{-\sqrt{-2\varepsilon^{(n)}}|\mathbf{r}-\mathbf{r}^\prime|}}{2\pi|\mathbf{r}-\mathbf{r}^\prime|}d\mathbf{r}^\prime,
    \label{eqn:Yukawa}
\end{equation}
and the Hartree potential,
\begin{equation}
    V_{H}(\mathbf{r}) = \int \frac{\rho(\mathbf{r}^\prime)}{|\mathbf{r}-\mathbf{r}^\prime|}d\mathbf{r}^\prime.
    \label{eqn:Coulomb}
\end{equation}
The singular ${\bf r}^\prime = {\bf r}$ term in the quadrature sums is skipped.
This error due to skipping the singularity is reduced by weakening the singularities before discretization.
For the integral involving the Yukawa kernel in Eq.~\eqref{eqn:Yukawa}
we implemented a standard singularity subtraction 
scheme~\cite{rabinowitz-1961, anselone-1981},
\begin{equation}
    \label{eqn:yukawa-conv-ss}
    \int  f(\mathbf{r}^\prime) \frac{e^{-k|\mathbf{r}-\mathbf{r}^\prime|}}{|\mathbf{r}-\mathbf{r}^\prime|}d\mathbf{r}^\prime =  \int \left(f(\mathbf{r}^\prime) - f(\mathbf{r})\right) \frac{e^{-k|\mathbf{r}-\mathbf{r}^\prime|}}{|\mathbf{r}-\mathbf{r}^\prime|}d\mathbf{r}^\prime + 
    f(\mathbf{r})\int \frac{e^{-k|\mathbf{r}-\mathbf{r}^\prime|}}{|\mathbf{r}-\mathbf{r}^\prime|}d\mathbf{r}^\prime.
\end{equation}
The second term on the right in Eq.~\eqref{eqn:yukawa-conv-ss} is evaluated analytically,
\begin{equation}
f(\mathbf{r})\int  \frac{e^{-k|\mathbf{r}-\mathbf{r}^\prime|}}{|\mathbf{r}-\mathbf{r}^\prime|}d\mathbf{r}^\prime = \frac{4\pi f(\mathbf{r})}{k^2},
\end{equation}
while the singularity in the first term on the right has been weakened,
so the quadrature scheme yields a more accurate result.
Note however that the exponential decay rate in the Yukawa kernel is
$k=\sqrt{-2\varepsilon}$, 
and 
a problem arises if $\varepsilon \to 0$, 
since in that case the singularity subtraction scheme in Eq.~\eqref{eqn:yukawa-conv-ss}
tends to the indeterminate form $\infty - \infty$.  
This is resolved by introducing a constant gauge shift in the effective potential, 
\begin{equation}
V_{eff}(\mathbf{r}) \to 
V_{eff}(\mathbf{r}) + V_{shift}.
\label{eqn:gauge-shift}
\end{equation}
The wavefunctions are unaffected 
and 
the eigenvalues simply shift by this amount
(the shift is removed before computing energies).  
Throughout this work we set $V_{shift} = -0.5$, 
ensuring that the eigenvalues of the occupied states are bounded away from zero.

The scheme described above however does not work for the 
Hartree potential in Eq.~\eqref{eqn:Coulomb}
which corresponds $k=0$.
In this case, we employ a modified form of singularity subtraction
using a Gaussian function,  
\begin{equation}
    \label{eqn:coulomb-conv-ss}
    \int  f(\mathbf{r}^\prime) \frac{1}{|\mathbf{r}-\mathbf{r}^\prime|}d\mathbf{r}^\prime = \int  \left(f(\mathbf{r}^\prime) - f(\mathbf{r})e^{-|\mathbf{r}-\mathbf{r}^\prime|^2/\alpha^2}\right) \frac{1}{|\mathbf{r}-\mathbf{r}^\prime|}d\mathbf{r}^\prime +
    f(\mathbf{r})\int  \frac{e^{-|\mathbf{r}-\mathbf{r}^\prime|^2/\alpha^2}}{|\mathbf{r}-\mathbf{r}^\prime|}d\mathbf{r}^\prime,
\end{equation}
where $\alpha$ is a scaling parameter.
As before, the second term on the right is evaluated analytically,
\begin{equation}
\label{eqn:correction_term}
f(\mathbf{r})\int  \frac{e^{-|\mathbf{r}-\mathbf{r}^\prime|^2/\alpha^2}}{|\mathbf{r}-\mathbf{r}^\prime|}d\mathbf{r}^\prime = 2\pi f(\mathbf{r})\alpha^2,
\end{equation}
and the singularity in the first term has been weakened.
In addition,
the Gaussian remains smooth for ${\bf r}^\prime \to {\bf r}$,
unlike other options,
and
this ensures the accuracy of the quadrature scheme.

The choice of the scaling factor $\alpha$ is guided by the
following considerations.
Recall that the first integral on the right in Eq.~\eqref{eqn:coulomb-conv-ss}
is computed using a quadrature scheme on a truncated computational domain,
and
hence the function $f(\mathbf{r})e^{-|\mathbf{r}-\mathbf{r}^\prime|^2/\alpha^2}$
should have certain properties.
If $\alpha$ is small, then the Gaussian is narrow 
and 
the quadrature scheme would struggle to resolve the variation in this function.  
On the other hand if $\alpha$ is large, then the Gaussian is wide
and
this would require increasing the size of the computational domain. 
In more detail,
the function $f(\mathbf{r})e^{-|\mathbf{r}-\mathbf{r}^\prime|^2/\alpha^2}$ 
must be sufficiently small when $\mathbf{r}^\prime$ is near the domain boundary
to ensure that the effect of the domain truncation is small;
there are two cases,
(1) when ${\bf r}$ lies in the domain interior,
then $f({\bf r})$ is not necessarily small,
but $e^{-|{\bf r}-{\bf r}^\prime|^2/\alpha^2}$ is small as long as
$\alpha$ is not too large,
(2) when ${\bf r}$ lies near the domain boundary,
then $f({\bf r})$ is small
while $e^{-|{\bf r}-{\bf r}^\prime|^2/ \alpha^2}$ is bounded.
The conclusion is that the Gaussian scaling factor $\alpha$
should not be too small in relation to the spatial discretization
and
should not be too large in relation to the computational domain size;
this work uses domains of size $[-20,20]^3$ a.u. to $[-30,30]^3$ a.u. 
with $\alpha = 1$ a.u., which was determined empirically.


\subsection{Gradient-Free Eigenvalue Update}
\label{section:gradient-free} 

Recall that line 4 in Green Iteration updates the  eigenvalue $\varepsilon_i^{(n+1)}$;
in this subsection we describe three methods for this purpose.
The first method uses the Rayleigh quotient~\cite{mohlenkamp-young-2008}, 
\begin{equation}
\varepsilon_i^{(n+1)} =  
\frac{ \langle \psi_i^{(n+1)}, \mathcal{H}\psi_i^{(n+1)} \rangle }{\langle \psi_i^{(n+1)},\psi_i^{(n+1)}\rangle} = \frac{
-\tfrac{1}{2}\langle \psi_i^{(n+1)},\nabla^2 \psi_i^{(n+1)} \rangle +
\langle \psi_i^{(n+1)},V_{eff} \psi_i^{(n+1)} \rangle}{\langle \psi_i^{(n+1)},\psi_i^{(n+1)}\rangle},
\label{eqn:laplacian-update}
\end{equation} 
where $\mathcal{H}$ is the Kohn-Sham differential operator 
defined in Eq.~\eqref{eqn:Kohn-Sham-differential},
$V_{eff}$ is the effective potential in the current SCF,
and
$\psi_i^{(n+1)}$ is the wavefunction computed in line 3 of Green Iteration.
The second method applies integration by parts in Eq.~\eqref{eqn:laplacian-update}
to obtain,
\begin{equation}
\varepsilon_i^{(n+1)} = \frac{ 
\tfrac{1}{2}\langle \nabla\psi_i^{(n+1)},\nabla\psi_i^{(n+1)} \rangle +
\langle \psi_i^{(n+1)},V_{eff} \psi_i^{(n+1)} \rangle}{\langle \psi_i^{(n+1)},\psi_i^{(n+1)}\rangle}.
\label{eqn:gradient-update}
\end{equation} 
In the present framework the gradient $\nabla \psi_i^{(n+1)}$ in Eq.~\eqref{eqn:gradient-update}
and
Laplacian $\nabla^2 \psi_i^{(n+1)}$ in Eq.~\eqref{eqn:laplacian-update}
are computed by spectral differentiation 
using the values of the wavefunction at the Chebyshev points in each cell~\cite{trefethen-2008}.
The third method is a gradient-free update suggested by Harrison et al. \cite{harrison-2004},
\begin{equation}
\label{eqn:gradient-free} 
\varepsilon_i^{(n+1)} = 
\varepsilon_i^{(n)} -
\frac{\langle V_{eff}\psi_i^{(n)},\psi_i^{(n)}-\psi_i^{(n+1)}\rangle}{\langle \psi_i^{(n+1)},\psi_i^{(n+1)}\rangle}.
\end{equation}
In this case,
which computes a $\Delta\varepsilon_i$ rather than $\varepsilon_i^{(n+1)}$ itself,
the initial guess $\varepsilon_i^{(0)}$ in the first SCF iteration 
can be given using either Eq.~\eqref{eqn:laplacian-update} or
Eq.~\eqref{eqn:gradient-update}.
The gradient-free eigenvalue update enables the 
total energy to also be computed in a gradient-free manner using the
alternative expression,
\begin{equation}
E = E_{band} - E_{H} + E_{xc} - \int\rho(\mathbf{r})V_{xc}[\rho]({\bf r})d\mathbf{r} + E_{ZZ},
\label{eqn:total-energy-gradient-free}
\end{equation}
where the band energy is the weighted sum of the eigenvalues,
\begin{equation}
E_{band} = 2\sum^{N_w}_{i=1}f_i(\varepsilon_i,\mu)\varepsilon_i.
\end{equation}
In contrast to the original expression for the 
total energy in Eq.~\eqref{eqn:total-energy-gradient}, 
the gradient-free expression in Eq.~\eqref{eqn:total-energy-gradient-free} 
avoids explicitly computing the kinetic energy $E_{kin}$ in Eq.~\eqref{eqn:kinetic-energy} 
which contains the Laplacian;  
the kinetic energy is now contained implicitly in the band energy, 
which is obtained with the gradient-free method.
Later below we show that the gradient-free method 
has the best accuracy of the three approaches described here.


\subsection{Accuracy Results} 
\label{section:Results-Accuracy}
This subsection demonstrates the effects of the 
previously described numerical techniques 
on the carbon monoxide molecule ($N_A=2$, $N_e=14$, $N_w=8$).  
The computations use domain $[-20,20]^3$ a.u.,
temperature $T=200~K$, 
gauge shift $V_{shift}=-0.5$, 
Green Iteration eigensolve tolerance $tol_{gi} = $1e$-7$, 
SCF tolerance $tol_{scf} = $1e$-6$, 
and Anderson mixing parameter $\beta=0.5$.
Except where specified, 
the computations use singularity subtraction and the gradient-free eigenvalue update. 
We report the energy error $|E_{TAGI} - E_{ref}|$,
where $E_{TAGI}$ is computed using TAGI
and
$E_{ref} = -112.47193$ Ha is the reference value converged to 1e$-4$ Ha,
which was computed using 
DFT-FE~\cite{motamarri-nowak-leiter-knap-gavini-2013,motamarri-das-rudraraju-ghosh-davydov-gavini-2019}.

\subsubsection{Quadrature Rule and Adaptive Mesh Refinement Scheme}
We first demonstrate the effect of the order $p$ of the quadrature rule 
and 
the tolerance $tol_m$ of the adaptive mesh refinement scheme described in section~\ref{section:quadrature}.
To test the effect of the quadrature rule order 
we generate a mesh using order $p=4$ and tolerance $tol_m$ = 3e$-7$, 
and
then on this mesh the order $p$ is varied;
table~\ref{table:mesh-refinement}a shows that 
the error is reduced from 1.313~mHa with $p=4$ to 0.179~mHa with $p=7$.
To test the effect of the mesh refinement tolerance 
we fix the quadrature order to $p=4$ and 
vary the mesh refinement tolerance $tol_m$;
table~\ref{table:mesh-refinement}b shows that 
the error is reduced 
from 3.946~mHa with $tol_m$ = 3e$-6$ 
to 0.674~mHa with $tol_m$ = 1e$-7$.
\begin{table}[htb]
\small
\centering
\begin{tabular}{ccccr|c}
(a) & $p$ & $tol_m$ & Cells & Points & Error (mHa) \\ \cline{2-6}
  & 4 & 3e-7 & 5293 & 661625 & 1.313 \\
  & 5 & 3e-7 & 5293 & 1143288 & 0.605 \\
  & 6 & 3e-7 & 5293 & 1815499 & 0.311 \\
  & 7 & 3e-7 & 5293 & 2710016 & 0.179 \\
\end{tabular}
\vfill
\vspace{0.5cm}
\begin{tabular}{ccccr|c}
(b) & $p$ & $tol_m$ & Cells & Points & Error (mHa) \\ \cline{2-6}
  & 4 & 3e-6 & 2962 & ~370250 & 3.946 \\
  & 4 & 1e-6 & 3676 & ~459500 & 2.550 \\
  & 4 & 3e-7 & 5293 & ~661625 & 1.313 \\
  & 4 & 1e-7 & 7428 & ~928500 & 0.674
\end{tabular}
\caption{
\textmd{
Error in the total energy per atom for the Carbon monoxide molecule using
(a) a fixed mesh, increasing quadrature order $p$ from 4 to 7, 
and
(b) a fixed quadrature order $p=4$, 
decreasing mesh refinement parameter $tol_m$ from 3e$-6$ to 1e$-7$.}
}
\label{table:mesh-refinement}
\end{table}


\subsubsection{Singularity Subtraction}

Next we demonstrate the effect of the singularity subtraction schemes described in section~\ref{section:singularity-subtraction}.  
The quadrature order is set to $p=4$ 
and 
a sequence of mesh refinements is performed.  
The ground state calculation is performed with and without singularity subtraction;
in both cases the singular term in the discrete convolution sums is skipped.
Table~\ref{table:singularity-subtraction} shows that 
singularity subtraction yields a significant improvement in the 
accuracy of the total energy,
over two orders of magnitude for mesh size $N_m = 661625$ 
which achieves chemical accuracy.
\begin{table}[htb]
\small
\centering
\begin{tabular}{ccc|c|c}
 & & & \multicolumn{2}{c}{Error (mHa)} \\\cline{4-5}
$tol_m$ & Cells & Points & 
Non-SS & 
SS\\ \hline
3e-6 & 2962 & 370250 & 823 & 3.946 \\
1e-6 & 3676 & 459500 & 702 & 2.550 \\
3e-7 & 5293 & 661625 & 558 & 1.313 \\
1e-7 & 7428 & 928500 & 429 & 0.674
\end{tabular}
\caption{
\textmd{
Error in the total energy per atom for the Carbon monoxide molecule 
without using singularity subtraction (column 4)
and with using singularity subtraction (column 5).}
}
\label{table:singularity-subtraction}
\end{table}

\subsubsection{Gradient-Free Eigenvalue Update}

\begin{figure}[htb]
\centering
\begin{subfigure}[b]{0.45\textwidth}
\centering
(a) order $p=4$
\includegraphics[width=\textwidth]{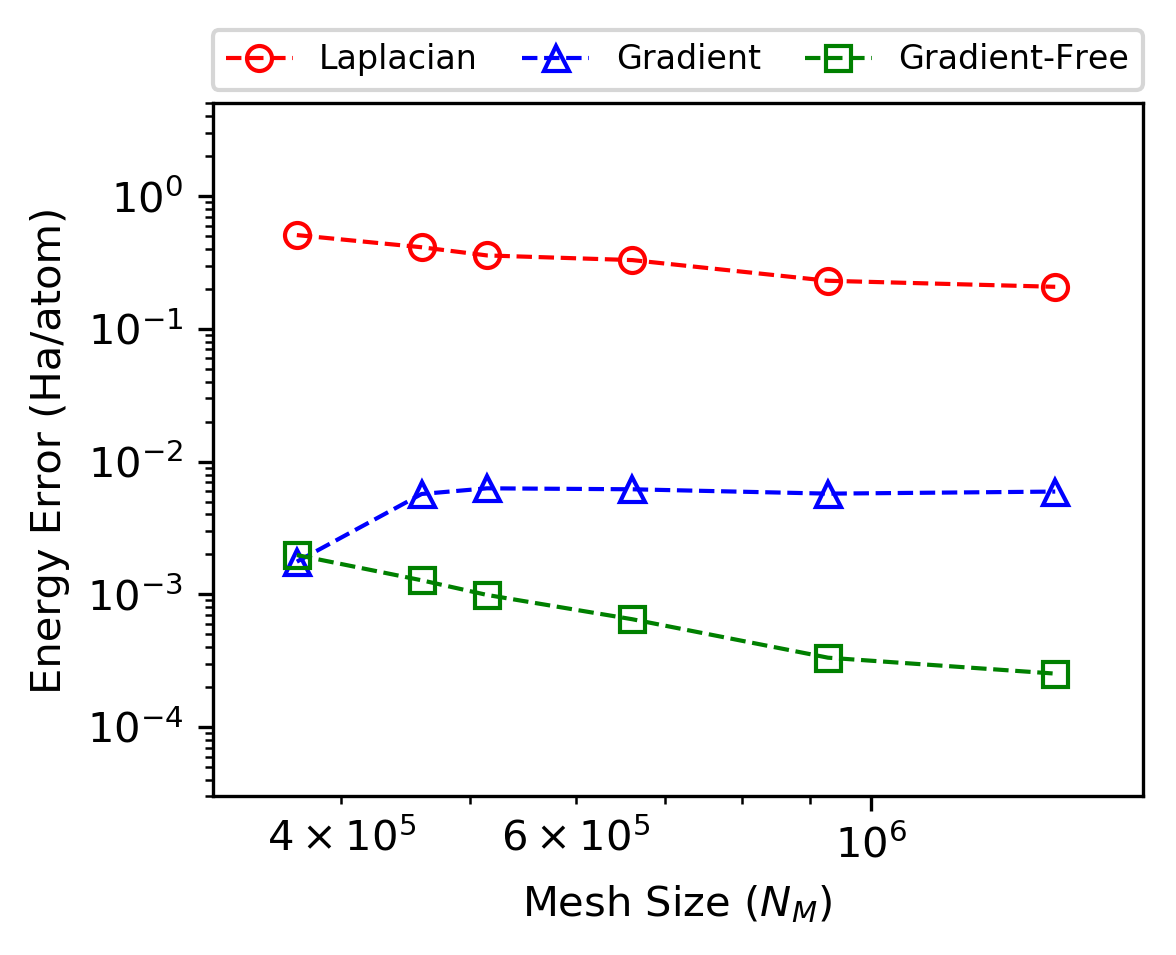}
\end{subfigure}
\begin{subfigure}[b]{0.45\textwidth}
\centering
(b) order $p=6$
\includegraphics[width=\textwidth]{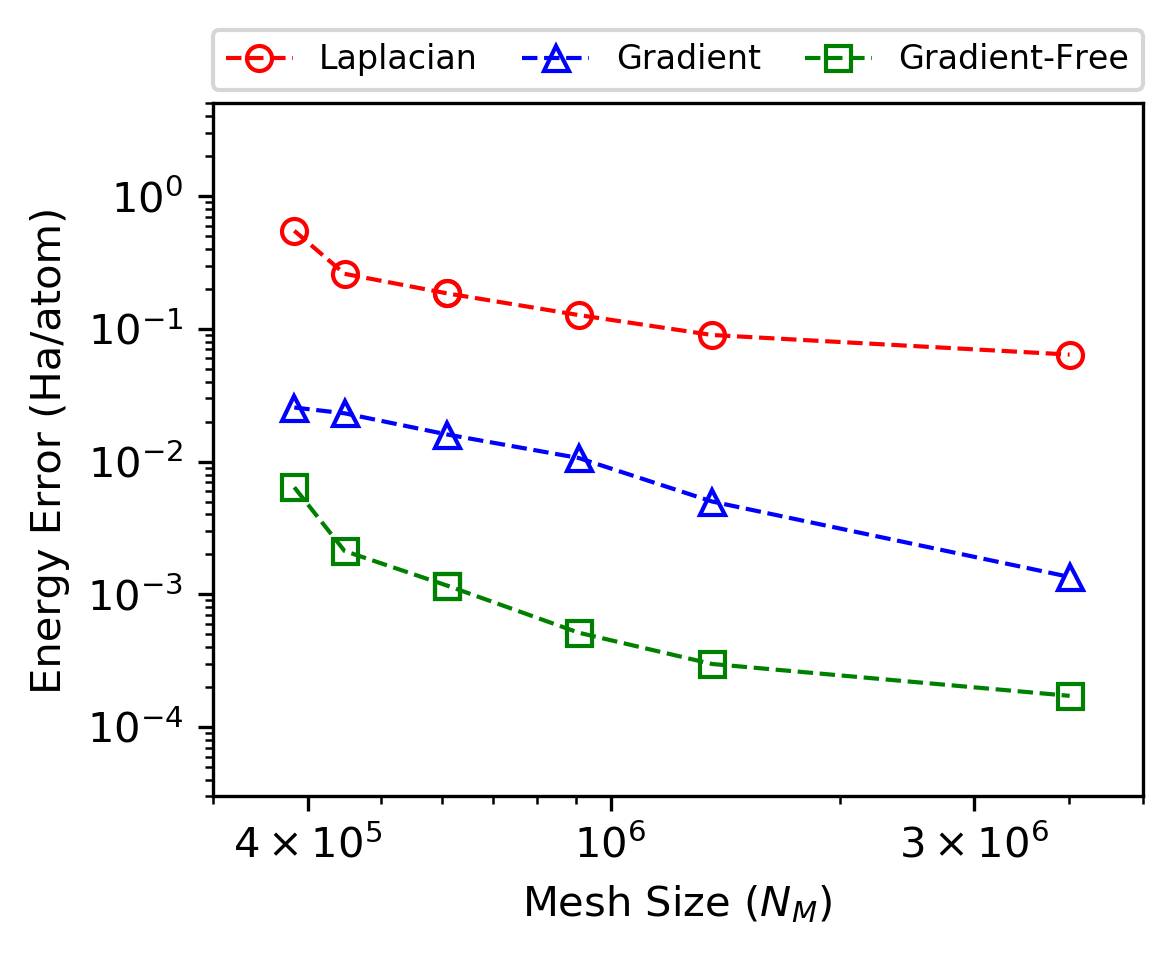}
\end{subfigure}
\caption{
\textmd{
Error in the total energy per atom for the Carbon monoxide molecule 
versus the number of mesh points $N_m$ while
using different eigenvalue update methods in Green Iteration for
(a) quadrature order $p=4$
and
(b) quadrature order $p=6$.}
}
\label{fig:gradient-free}
\end{figure}

Finally, we compare the eigenvalue update methods 
described in section~\ref{section:gradient-free}, 
Laplacian update (Eq.~\eqref{eqn:laplacian-update}), 
gradient update (Eq.~\eqref{eqn:gradient-update}), and 
gradient-free update (Eq.~\eqref{eqn:gradient-free}).
The ground state calculation is performed for a sequence of refined meshes.
Figure~\ref{fig:gradient-free}a shows the energy error versus 
the number of mesh points $N_m$ for order $p=4$
and
figure~\ref{fig:gradient-free}b shows this for order $p=6$.
We make the following three observations.
First, for a given mesh, the gradient-free update achieves significantly better accuracy than the gradient and Laplacian updates for both order $p=4$ and $p=6$,
and as the mesh is refined the gradient-free update achieves chemical accuracy around $N_m=600,000$.
Second, for $p=4$, the gradient update error saturates around 6 mHa/atom as the mesh is refined, 
indicating that the refinement scheme is not adequately refining the correct regions to reduce the error in the kinetic energy.
Third, for $p=6$, the gradient update recovers its convergence rate and is able to achieve chemical accuracy as the mesh is refined, indicating that the higher order gradients have reduced the error in the kinetic energy that was present for $p=4$.
We note that the adaptive mesh refinement scheme and choice of test function described in section~\ref{section:quadrature} were developed using feedback from the gradient-free eigenvalue update.
Different meshing schemes that prioritize accurate gradients or Laplacians of the wavefunctions could achieve better results for their respective eigenvalue update methods than this refinement scheme.
Nevertheless, for each mesh refinement scheme we investigated we found the gradient-free update to be the most accurate and we use this update throughout the work.


\section{Convergence Rate of Green Iteration} \label{section:wavefunction-mixing}

Previously we explained how in each SCF iteration,
the Kohn-Sham eigenproblem in Eq~\eqref{eqn:Kohn-Sham-differential}
can be converted into a fixed-point problem 
for the integral operator in Eq.~\eqref{eqn:Integral_Form}
and
that the fixed-point problem is solved by Green Iteration.
This section examines the convergence rate of Green Iteration;
first an example exhibiting slow convergence is presented, 
then the cause of the problem is identified by reference to power iteration,
and
finally Anderson mixing is applied to the wavefunctions
to accelerate convergence.

\subsection{Slow convergence of Green Iteration} \label{section:slow_conv_GI}

\begin{figure}[htb]
\centering
\begin{subfigure}[b]{\textwidth}
    \centering
\includegraphics[width=\textwidth]{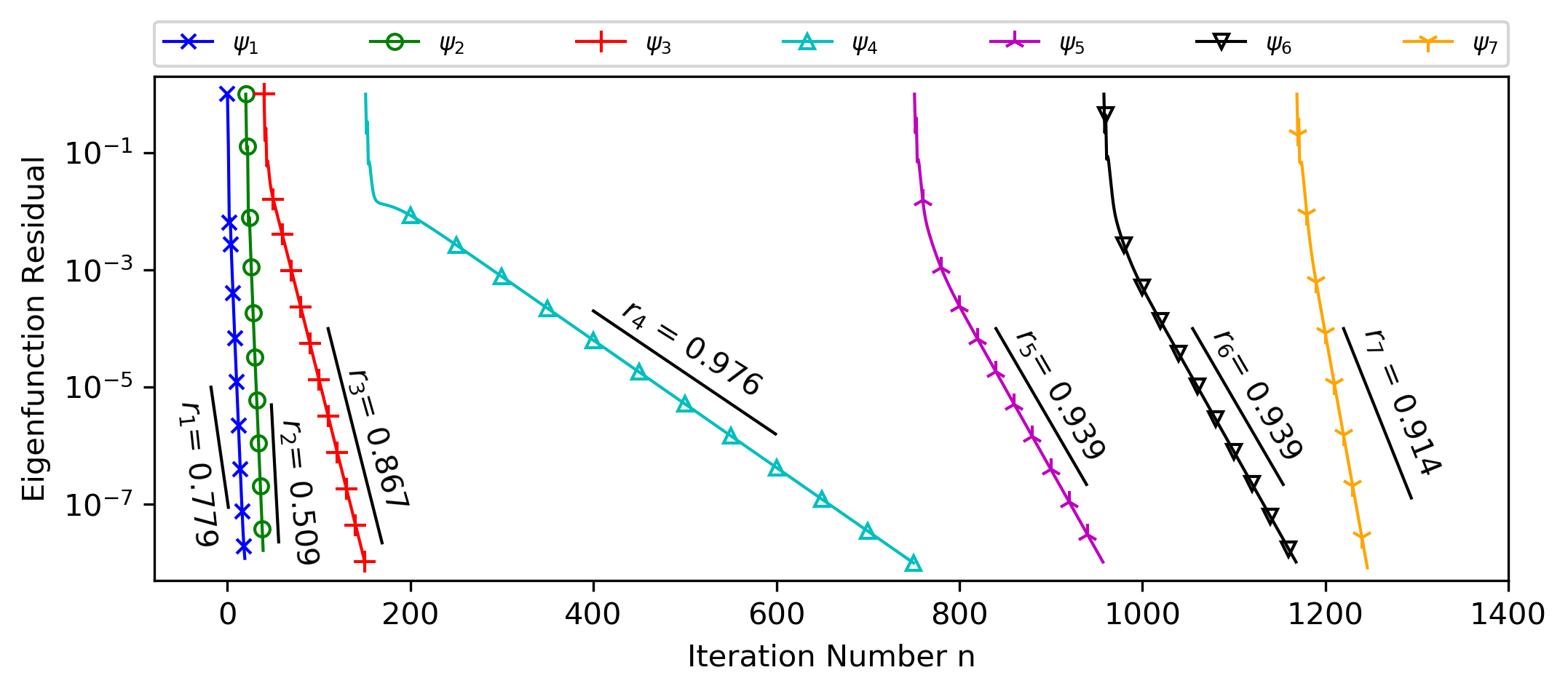}
\end{subfigure}
\caption{ 
\textmd{Convergence of the eigenfunction residual $||\psi_i^{(n+1)} - \psi_i^{(n)}||_2$ during Green Iteration in the first SCF iterations for the
carbon monoxide molecule.
The observed residuals (symbols) and the
predicted convergence rates $r_i$ (black lines).}
}
\label{fig:gi-convergence}
\end{figure}

To illustrate the slow convergence of Green Iteration,
we consider the first SCF iteration for the carbon monoxide molecule.
Figure~\ref{fig:gi-convergence} plots the residual of the 
first seven wavefunctions determined by Green Iteration versus
the iteration number.
In this case the first two wavefunctions converge rapidly,
but the subsequent wavefunctions converge slowly;
in particular the 4th wavefunction converges extremely slowly.
The result is that Green Iteration requires a total of 1246 iterations 
to ensure that the first seven wavefunction residuals fall below 1e$-$8.
This is a tighter tolerance than is used in practice, 
however it helps illustrate the issue.
In the next subsection we examine the cause of this slow convergence.

\subsection{Convergence Analysis}  
\label{section:GI-convergence} 

Recall line 3 of Green Iteration (Algorithm~\ref{alg:GreensIteration}),
$\psi_i^{(n+1)} = \mathcal{G}(\varepsilon_i^{(n)})\psi_i^{(n)}$,
which updates the $i$th eigenfunction using the
operator 
defined in Eq.~\eqref{eqn:integral_operator}.
The parameter $\varepsilon_i^{(n)}$ changes in each step of the iteration,
but as $\varepsilon_i^{(n)} \to \varepsilon_i$, 
the scheme converges to power iteration for the 
operator $\mathcal{G}(\varepsilon_i)$ 
with deflation against the previously determined eigenfunctions 
$\psi_j, j < i$ as
indicated in line 5 of the algorithm.
This suggests that the convergence rate of $\psi_i^{(n)}$ 
depends on the spectral gap of $\mathcal{G}(\varepsilon_i)$~\cite{trefethen-bau-2000}.
%
%
To demonstrate this it is useful to
define a 1-parameter family of curves $\mu_i(\varepsilon)$ and functions $\phi_i(\varepsilon)$
satisfying the linear eigenvalue equation,
\begin{equation}
\mathcal{G}(\varepsilon)\phi_i(\varepsilon) = \mu_i(\varepsilon)\phi_i(\varepsilon),
\quad i = 1, \ldots, N_w,
\label{eqn:mu-phi-equation}
\end{equation}
subject to the following conditions.
For each $\varepsilon$, the eigenvalues are ordered by their magnitude
$\mu_1(\varepsilon) \geq \cdots \geq \mu_{N_w}(\varepsilon)$.
Note that if $\mu_i(\varepsilon) = 1$ 
for some index $i$ and parameter value $\varepsilon$,
then Eq.~\eqref{eqn:mu-phi-equation} reduces to the fixed-point problem in Eq.~\eqref{eqn:Integral_Form},
$\psi_i = \mathcal{G}(\varepsilon_i)\psi_i$,
in which case we have 
$\varepsilon = \varepsilon_i$
and
$\phi_i(\varepsilon) = \psi_i$~\cite{mohlenkamp-young-2008,khoromskij-2008}.
In addition,
the usual orthogonality condition, 
$\phi_i(\varepsilon) \perp \phi_j(\varepsilon)$ for $i \ne j$, 
is modified to be consistent with the deflation step in Green Iteration;
that is, 
$\psi_i \perp \phi_j(\varepsilon)$ for $i < j$
and
$\varepsilon_i < \varepsilon$.





\begin{figure}[htb]
\centering
\begin{subfigure}[b]{0.32\textwidth}
\centering
\includegraphics[width=\textwidth]{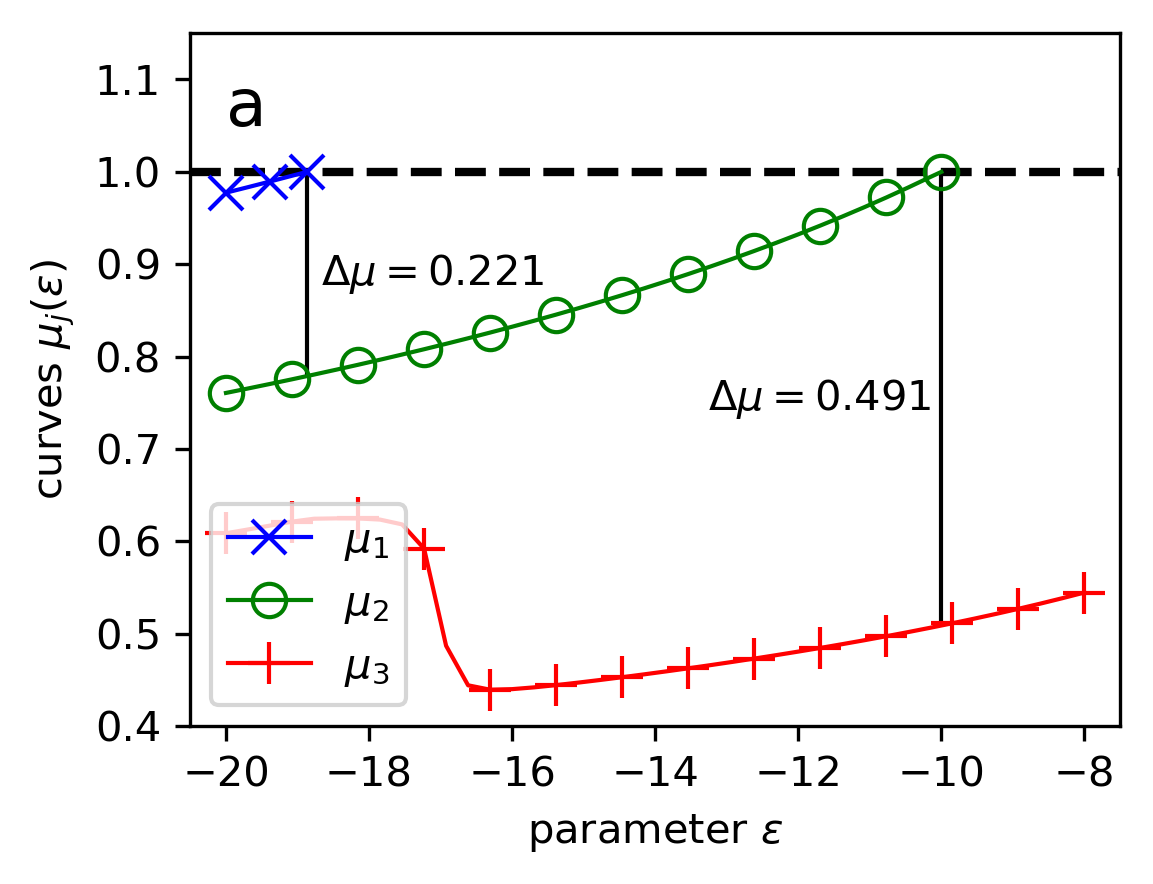}
\end{subfigure}
\begin{subfigure}[b]{0.32\textwidth}
\centering
\includegraphics[width=\textwidth]{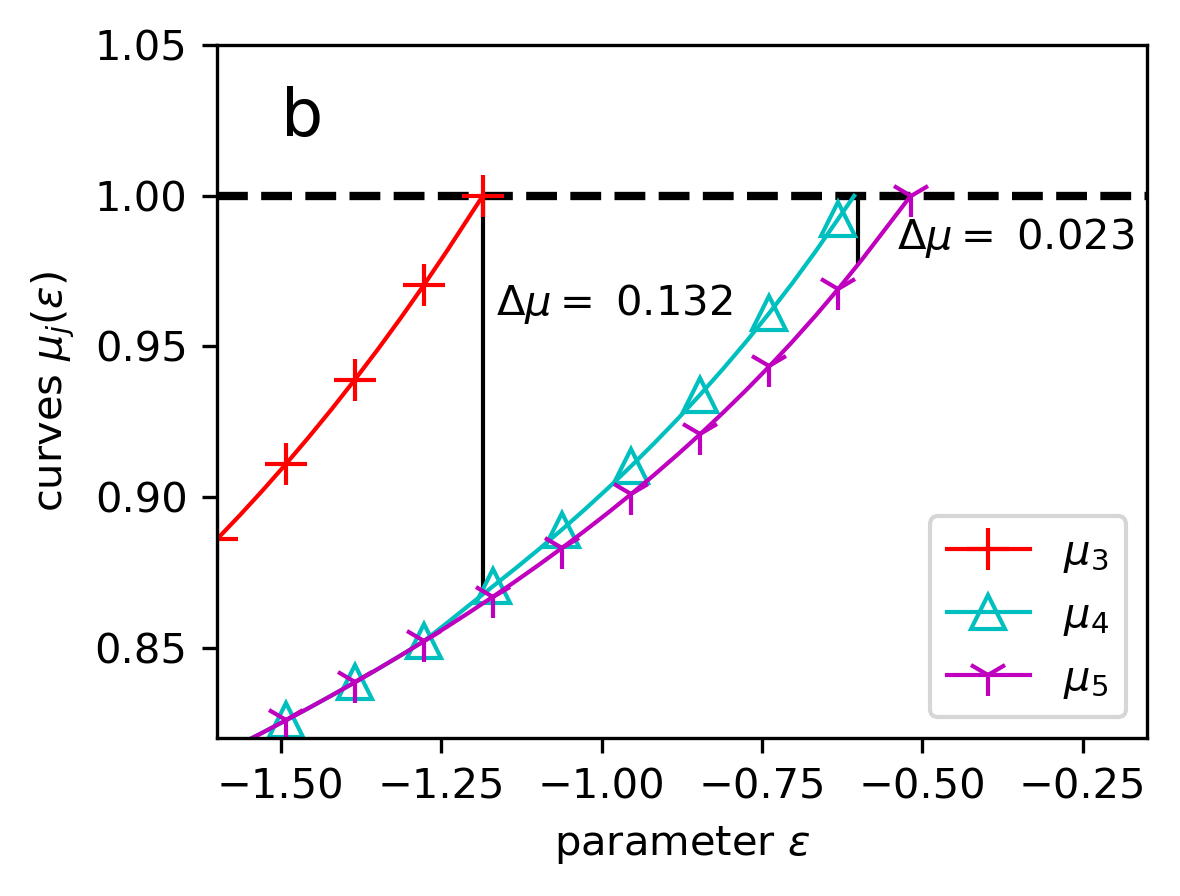}
\end{subfigure}
\begin{subfigure}[b]{0.32\textwidth}
\centering
\includegraphics[width=\textwidth]{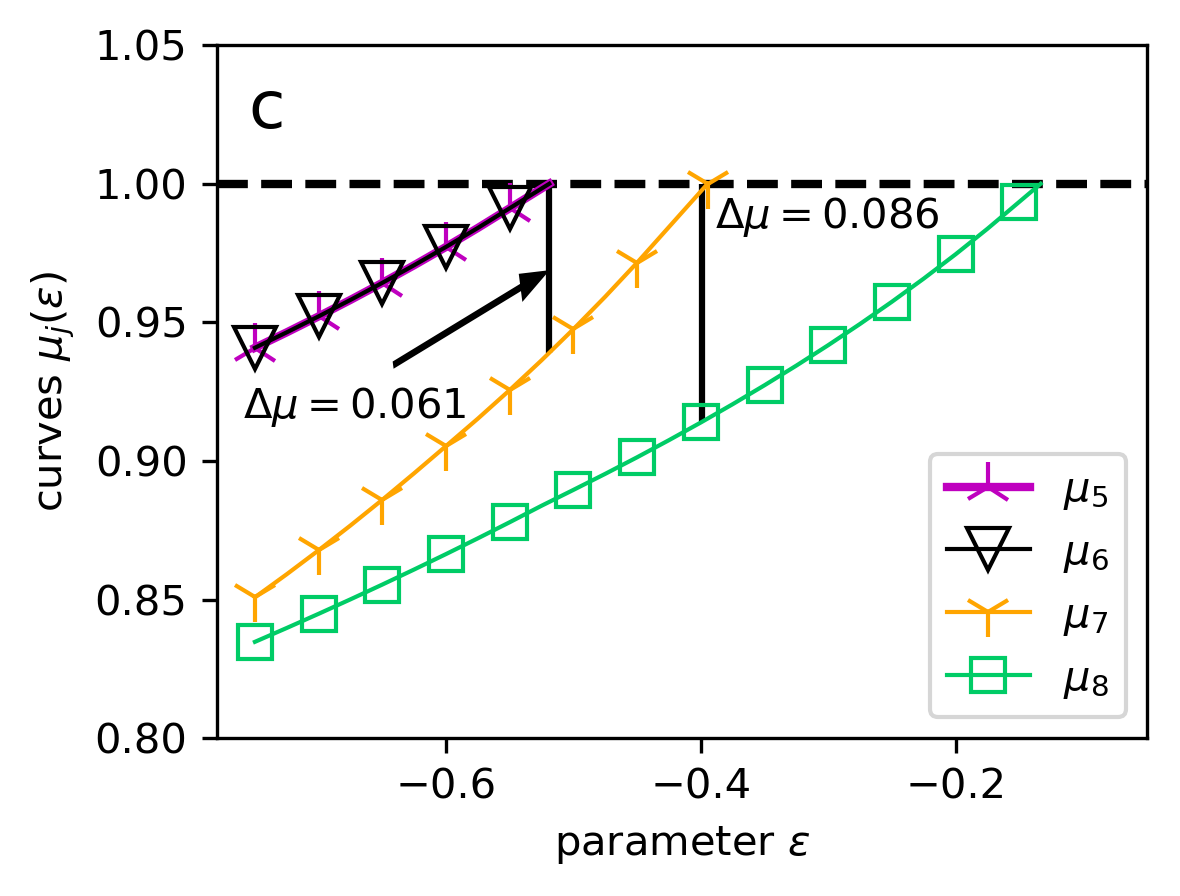}
\end{subfigure}
\caption{
\textmd{
First SCF iteration for the Carbon monoxide molecule. 
The curves $\mu_i(\varepsilon)$ defined by eigenvalue problem
Eq.~\eqref{eqn:mu-phi-equation} for the integral operator $\mathcal{G}(\varepsilon)$
are plotted versus parameter $\varepsilon$.
Intersections with the dashed line $\mu =1$ yield fixed-points $\varepsilon_i$ of Green Iteration.
The spectral gap of the integral operator,
$\Delta\mu_i = 1 - \mu_{i+1}(\varepsilon_i)$, is indicated 
at the fixed-points $\varepsilon_i$,
(a) $i=1,2,3$,
(b) $i=3,4,5$, 
(c) $i=5,6,7,8$.
Numerical values are given in Table~\ref{tab:gaps}.}
}
\label{fig:eigenvalue-one-sweep}
\end{figure}

Figure~\ref{fig:eigenvalue-one-sweep}
illustrates this for the first SCF iteration of the carbon monoxide molecule,
where the curves $\mu_i(\varepsilon)$ are plotted versus $\varepsilon$
for $i = 1:8$.
Note that for each parameter value $\varepsilon$,
the eigenvalues $\mu_i(\varepsilon)$ are computed by power iteration 
applied to the operator $\mathcal{G}(\varepsilon)$
subject to the modified orthogonality condition stated above.
The fixed-points of Green Iteration occur when one of the curves $\mu_i(\varepsilon)$
intersects the line $\mu=1$,
and
the plotted curves terminate there because there are no
eigenfunctions with $\mu > 1$ due to the orthogonality condition.


Figure~\ref{fig:eigenvalue-one-sweep}
also indicates the spectral gap of the operator $\mathcal{G}(\varepsilon_i)$,
defined by $\Delta\mu_i = 1 - \mu_{i+1}(\varepsilon_i)$;
due to the continuity of the curves $\mu_i(\varepsilon)$,
these are correlated with the
spectral gap of the Hamiltonian,
defined by
$\Delta\varepsilon_i = \varepsilon_{i+1} - \varepsilon_i$;
hence
both gaps are relatively large for $i = 1,2$ in Fig.~\ref{fig:eigenvalue-one-sweep}a,
and 
relatively small for $i = 3,...,7$ in Fig.~\ref{fig:eigenvalue-one-sweep}b,c.
Note further that the spectrum of the CO molecule contains a degeneracy;
$\varepsilon_5=\varepsilon_6$, hence $\psi_5$ and $\psi_6$ span a degenerate subspace. 
This degeneracy manifests itself in the spectral analysis in several ways.
First, $\psi_5$ and $\psi_6$ converge with identical rates in Green Iteration (Fig.~\ref{fig:gi-convergence} parallel purple and black),
and second, 
the $\mu_5(\varepsilon)$ and $\mu_6(\varepsilon)$ curves are identical (Fig.~\ref{fig:eigenvalue-one-sweep}c overlapping purple and black).
In the case of a degeneracy, the convergence rate of the wavefunctions to the degenerate subspace is governed by the spectral gap to the next distinct eigenvalue.  
In this example, we define the spectral gaps $\Delta\mu_5$ and $\Delta\mu_6$ with respect to the 7th eigenvalue,
$\Delta\mu_5 = 1 - \mu_{7}(\varepsilon_5)$ and
$\Delta\mu_6 = 1 - \mu_{7}(\varepsilon_6)$.
Finally, note that at a fixed-point parameter $\varepsilon_i$,
the largest eigenvalue of $\mathcal{G}(\varepsilon_i)$ is $\mu_i=1$, 
so the convergence rate of power iteration is $r_i = \mu_{i+1}/\mu_i = 1-\Delta\mu_i$;
hence a large gap $\Delta\mu_i$ leads to rapid convergence of $\psi_i$
and
a small gap $\Delta\mu_i$ leads to slow convergence.




Table~\ref{tab:gaps} gives the values of the fixed-points, the spectral gaps, the observed and predicted convergence rates, 
the accuracy of the predictions, and the number of iterations required to achieve the 1e-8 tolerance in Green Iteration. 
The predicted convergence rates $r_i$, 
also shown in Fig.~\ref{fig:gi-convergence},
were computed using the power iteration considerations above,
$r_i = \mu_{i+1}/\mu_i = 1-\Delta\mu_i$.
In several cases ($\psi_1,~\psi_2,~\psi_7$), the observed convergence is faster than the predicted rate;
this is attributed to the iteration not entering the asymptotic power-iteration regime before the tolerance was met.
In the slower converging cases ($\psi_3,~\psi_4,~\psi_5,~\psi_6$), the predicted convergence rates accurately agree with the observed rates,
with percent errors $0.115\%$, $0.082\%$, $0.053\%$, and $0.053\%$,
confirming that the convergence rates of the eigenfunctions $\psi_i$ in Green Iteration
are controlled by the spectral gap in the integral operator $\Delta\mu_i$, 
which are correlated to the spectral gap in the differential operator $\Delta\varepsilon_i$.
Hence,
Green Iteration may converge slowly whenever a small spectral gap exists in the Hamiltonian;
the next subsection describes a method to overcome this drawback.
\begin{table}[htb]
    \centering
    \begin{tabular}{c|ccc|ccc|c}
 & \multicolumn{3}{c|} {Spectral Gaps} & \multicolumn{3}{c|} {Convergence Rates} & \\
index, $i$ & $\varepsilon_i$ & $\Delta\varepsilon_i$ & $\Delta\mu_i$ & observed $r_i$ & predicted $r_i$ & $\%$ error & Number of Iterations \\
\hline
1 & -18.870 & 8.862 & 0.221 & 0.460  & 0.779 & 69.3 & 19\\
2 & -10.008 & 8.822 & 0.491 & 0.450  & 0.509 & 13.1 & 20\\
3 & -1.186  & 0.579 & 0.132 & 0.867  & 0.868  & 0.115 & 110\\
4 & -0.607  & 0.087 & 0.023 & 0.9752 & 0.976 & 0.082 & 601\\
5 & -0.520  & 0.124 & 0.061 & 0.9385 & 0.939 & 0.053 & 208\\
6 & -0.520  & 0.124 & 0.061 & 0.9385 & 0.939 & 0.053 & 211\\
7 & -0.396  & 0.262 & 0.086 & 0.819  & 0.914  & 11.6 & 77\\
\end{tabular}
\caption{
\textmd{First SCF iteration for the Carbon monoxide molecule. 
Eigenvalue index, 
Hamiltonian eigenvalues $\varepsilon_i$, Hamiltonian spectral gap $\Delta\varepsilon_i$, 
integral operator spectral gap $\Delta\mu_i$, 
observed convergence rate,
predicted convergence rate, 
accuracy of the prediction,
number of iterations for the wavefunction to converge to 1e-8 tolerance.}
}
\label{tab:gaps}
\end{table}

\subsection{Wavefunction Mixing}  
\label{section:anderson-GI} 

\begin{figure}[htb]
\centering
\begin{subfigure}[b]{0.8\textwidth}
    \centering
\includegraphics[width=\textwidth]{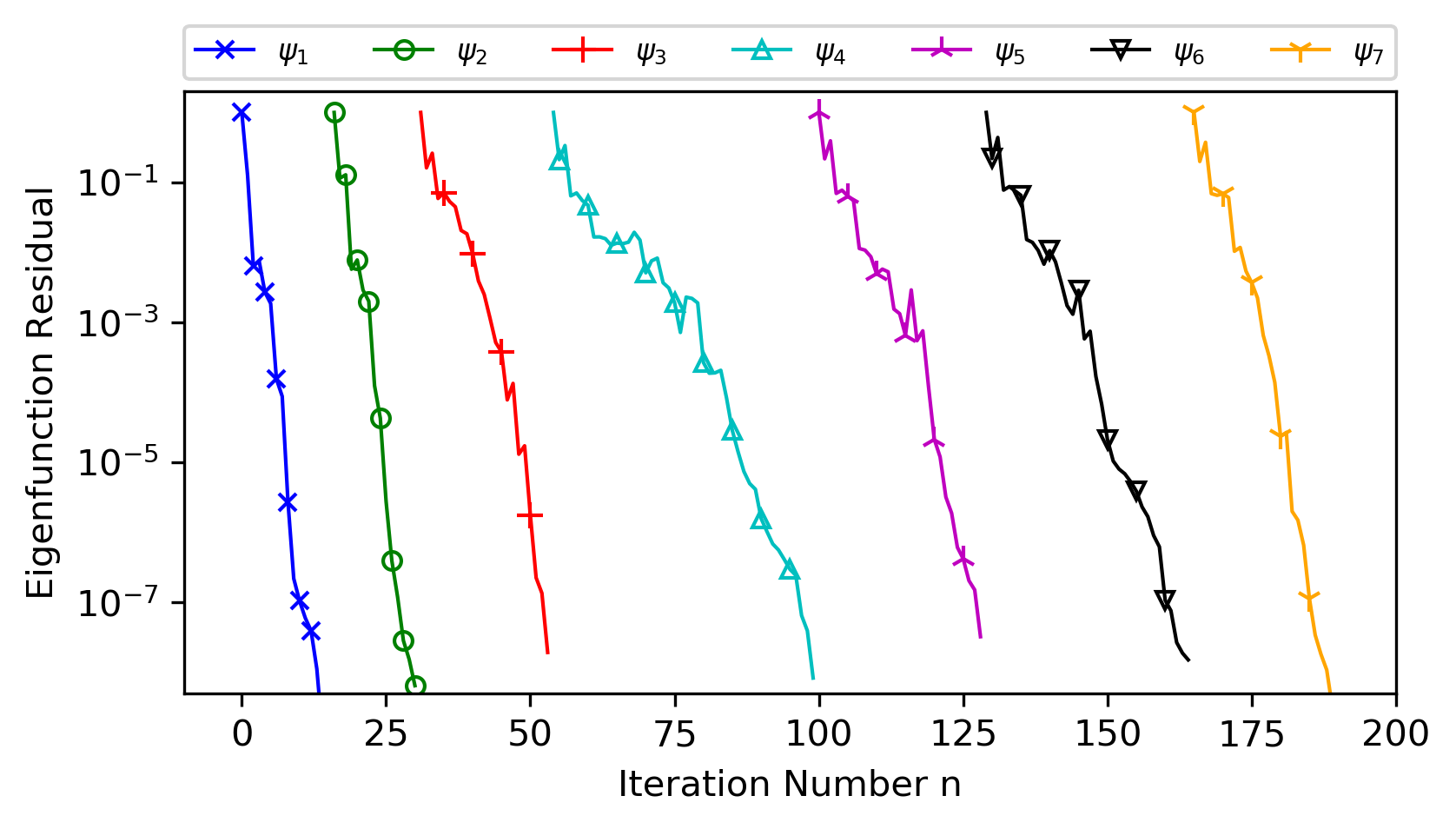}
\end{subfigure}
\caption{
\textmd{
Convergence of the eigenfunction residual $||\psi_i^{(n+1)} - \psi_i^{(n)}||_2$ during Green Iteration in the first SCF iterations for the
carbon monoxide molecule using wavefunction mixing with mixing parameter $\beta = 0.5$.
Wavefunction mixing reduces the total number of iterations from 1246 to 188.}
}
\label{fig:wavefunction-mixing}
\end{figure}

While Green Iteration resembles power iteration as noted above, 
it is a fixed-point iteration 
and 
hence is amenable to standard fixed-point acceleration techniques.  
We define the vector $\mathbf{x} = (\varepsilon,\psi)$, 
and the inner product between two vectors 
$\mathbf{x}_1=(\varepsilon_1, \psi_1)$ and $\mathbf{x}_2=(\varepsilon_2, \psi_2)$ 
to be 
$(\mathbf{x}_1,\mathbf{x}_2) = \varepsilon_1\varepsilon_2 + \int \psi_1({\bf r})\psi_2({\bf r})d{\bf r}$.
We then use Anderson mixing to update the eigenpairs $(\varepsilon_i^{(n)},\psi_i^{(n)})$
after each step of Green Iteration, 
in the same way that the electron density is updated after each step of the SCF iteration.
Figure~\ref{fig:wavefunction-mixing} shows the effect of applying
Anderson mixing to the wavefunctions with mixing parameter $\beta = 0.5$,
for the same computation as above, 
the first SCF iteration of the carbon monoxide molecule. 
The total number of iterations is reduced from 1246 (Green Iteration) 
to 188 (Green Iteration with wavefunction mixing).  
$\psi_3 - \psi_6$ still converge the slowest, however they converge significantly faster than without Anderson mixing.

In practice
the wavefunction mixing scheme requires a good initial guess to ensure convergence.
In the first SCF iteration, 
to achieve a good initial guess,
Green Iteration can be performed without wavefunction mixing until convergence to a 
user-defined tolerance is achieved, 
at which point the computed wavefunction is in the basin of attraction of the fixed-point scheme
and Anderson wavefunction mixing can be safely applied.  
In subsequent SCF iterations the initial guess for the eigenpairs tend to be much better 
and 
delaying the use of wavefunction mixing is not necessary.
Furthermore,
the tolerance for Green Iteration $tol_{gi}$ does not have to be the same throughout an SCF iteration.
We find that starting with a loose tolerance and gradually tightening it after each step in the SCF
is beneficial.
The gradual reduction of $tol_{gi}$ increases the number of steps in the SCF for the 
electron density to converge to $tol_{scf}$,
but it significantly reduces the cost of the first few steps of the SCF iteration, 
and results in an overall reduction of computation time.


\section{Treecode Acceleration} \label{section:treecode}

The scheme described above requires computing convolutions of the form, 
\begin{equation}
\label{eqn:convolution_integral}
u(\mathbf{r}) = 
\int G(\mathbf{r},\mathbf{r}^\prime)f(\mathbf{r}^\prime)d\mathbf{r}^\prime,
\end{equation}
where $G({\bf r},{\bf r}^\prime)$ 
is either the Coulomb kernel in the Hartree potential in Eq.~\eqref{eqn:classical-electrostatics},
or the Yukawa kernel in the integral operator $\mathcal{G}(\varepsilon)$ 
in Eq.~\eqref{eqn:integral_operator}
needed in line 3 of Green Iteration.
Upon discretization,
the integral in Eq.~\eqref{eqn:convolution_integral} is approximated by the discrete convolution sum,
\begin{equation}
u_i = \sum_{\substack{j=1 \\ i\neq j}}^{N_m} G(\mathbf{r}_i,\mathbf{r}_j)f_j w_j, \quad 
i=1, \ldots, N_m,
\label{eqn:direct-sum}
\end{equation}
where 
$u_i \approx u({\bf r}_i)$,
$f_j = f({\bf r}_j)$,
and
$w_j$ are the quadrature weights.
Computing $u_i$ by direct summation requires $O(N_m^2)$ operations,
and
several methods have been developed to reduce the cost including
the treecode~\cite{barnes-1986} and fast multipole method~\cite{greengard-rokhlin-1987}.
This work employs a recently developed
barycentric Lagrange treecode (BLTC)~\cite{wang-krasny-tlupova-2019}
which reduces the operation count to $O(N_m\log N_m)$
using barycentric Lagrange interpolation~\cite{berrut-trefethen-2004}.
For clarity of presentation, the singularity subtraction schemes from equations~\eqref{eqn:yukawa-conv-ss} and~\eqref{eqn:coulomb-conv-ss} have been omitted in Eq.~\eqref{eqn:direct-sum},
but they are easily accommodated in the BLTC and are used in practice.
Following convention,
throughout this section the points $\mathbf{r}_i$ are referred to as target particles,
the points $\mathbf{r}_j$ are referred to as source particles,
and
Eq.~\eqref{eqn:direct-sum} expresses the particle-particle interactions.
Below we present an overview of the treecode;
references can be consulted for more details~\cite{wang-krasny-tlupova-2019,vaughn-wilson-krasny-2020,li-2009}.


\subsection{Source Clusters and Target Batches}

The treecode starts by dividing the source particles into a 
hierarchical tree of source clusters,
where the root cluster is the minimal bounding box enclosing the computational domain.
The root is divided into child clusters by bisection in each dimension,
and
the child clusters are recursively subdivided until they contain fewer than $N_L$ particles;
these are the leaves of the tree.
After division each cluster is shrunk to the minimal bounding box containing its particles. 
Typically a cluster is divided into eight children, 
but if shrinking would cause the aspect ratio to be greater than $\sqrt{2}$,
the cluster is instead divided into either two or four children.
Note that the source clusters in the treecode are rectangular boxes,
and
in general they are different than the cells in the adaptive mesh.
For efficiency purposes as explained below,
the target particles are also organized into a set of localized batches containing fewer than $N_B$ particles, 
and then the particle-particle interactions are organized into batch-cluster interactions
between the target particles in a batch and the source particles in a cluster.
In this work we set $N_B=N_L$, 
and
since the target particles and source particles correspond to the same set (the $N_m$ quadrature points),
the target batches are equivalent to the leaf source clusters in the tree.


\subsection{Particle-Cluster Approximation by Barycentric Lagrange Interpolation}
\label{section:particle-cluster-interactions}  

Note that the sum in Eq.~\eqref{eqn:direct-sum} can be rewritten as
\begin{equation}
u_i = 
\sum_{\substack{j=1 \\ i\neq j}}^{N_m} G(\mathbf{r}_i,\mathbf{r}_j)f_j w_j =
\sum_C u({\bf r}_i,C),
\end{equation}
where the second sum is taken over a set of source clusters $C$,
and
\begin{equation}
u({\bf r}_i,C) =
\sum_{{\bf r}_j \in C} G(\mathbf{r}_i,\mathbf{r}_j)f_j w_j
\label{eqn:direct-sum_2}
\end{equation}
is the interaction between a target particle ${\bf r}_i$ 
and
a source cluster $C = \{{\bf r}_j\}$.
Following~\cite{wang-krasny-tlupova-2019},
the particle-cluster interaction can be approximated by 
3D polynomial interpolation of degree $n$,
\begin{equation}
u(\mathbf{r}_i,C) \approx  
\sum_{k_1=0}^{n} \sum_{k_2=0}^{n} \sum_{k_3=0}^{n} 
G(\mathbf{r}_i,\mathbf{s_k})\widehat{f}_{\bf k}, \quad
\widehat{f}_{\bf k} = 
\sum_{\mathbf{r}_j\in C}L_{k_1}(r_{j1})L_{k_2}(r_{j2})L_{k_3}(r_{j3})f_jw_j,
\label{eqn:particle-cluster-approx-rearranged}
\end{equation}
where ${\bf r}_j = (r_{j1},r_{j2},r_{j3})$ is a source particle,
${\bf s}_{\bf k} = (s_{k_1},s_{k_2},s_{k_3})$ is a tensor product grid of interpolation points,
$L_k(t)$ are the 1D Lagrange interpolating polynomials,
and
$\widehat{f}_{\bf k}$ are weights associated with the approximation.
It is important to note that the
approximation in Eq.~\eqref{eqn:particle-cluster-approx-rearranged}
has the same direct sum structure as the 
exact interaction in Eq.~\eqref{eqn:direct-sum_2};
in one case the target particle ${\bf r}_i$ interacts with the
source particles ${\bf r}_j$
and 
in the other it interacts with the interpolation points ${\bf s}_k$;
however in both cases the necessary kernel evaluations are independent of each other
and can be efficiently computed in parallel on a GPU;
this is in contrast to other fast summation schemes based on analytic series expansions,
such as the Taylor treecode~\cite{li-2009},
where the approximations are recursive, which limits performance on the GPU.
A further point is that the approximation weights $\widehat{f}_{\bf k}$
are independent of the target particle ${\bf r}_i$,
so they can be precomputed and reused for different targets.
Next we examine the decision of when to use the approximation in Eq.~\eqref{eqn:particle-cluster-approx-rearranged}, 
then return to the choice of interpolation points ${\bf s}_{\bf k}$ and the structure of the interpolating polynomial $L_k(t)$.

\begin{figure}[htb]
\centering
\includegraphics[width=0.6\textwidth]{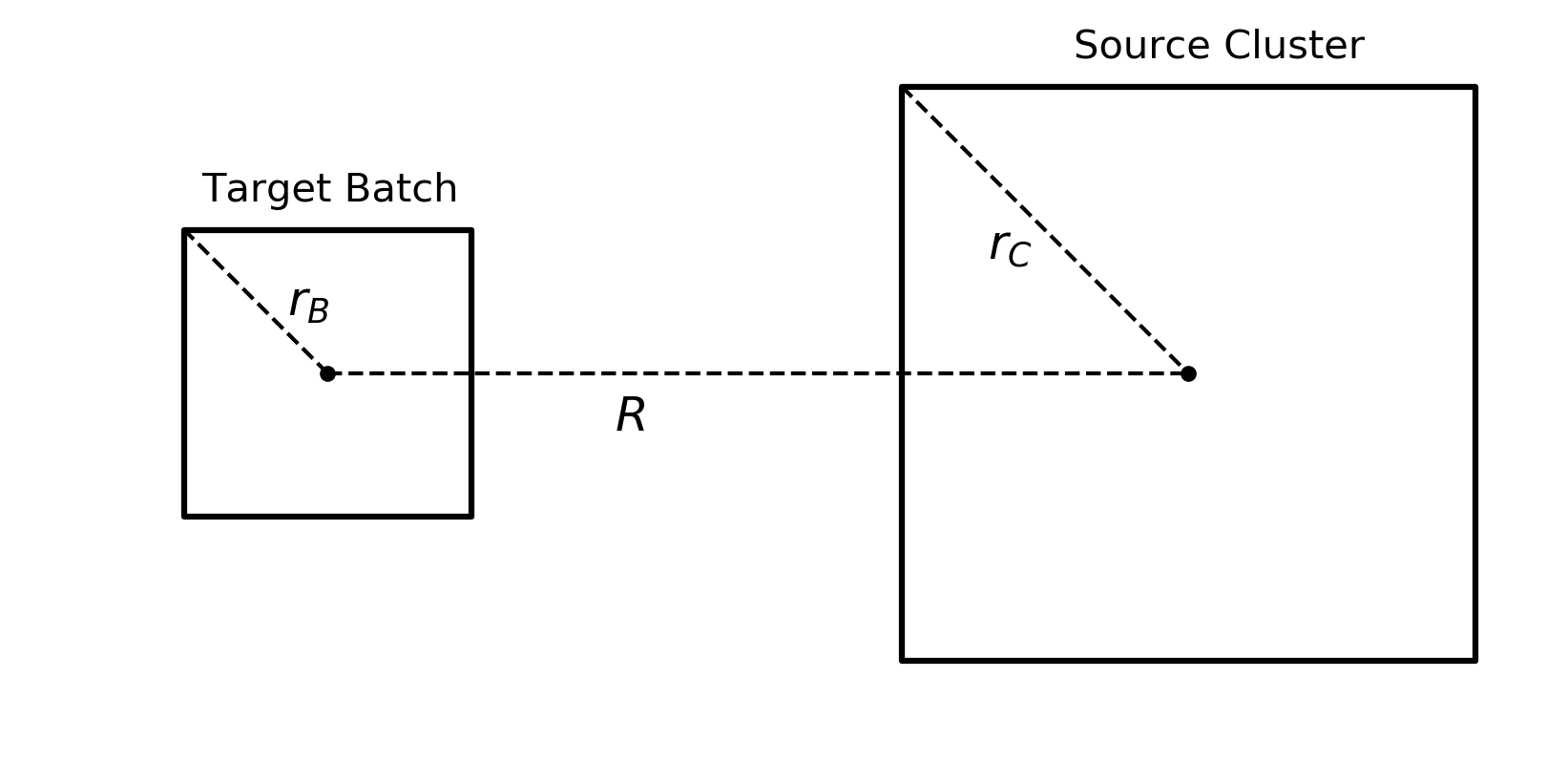}
\caption{
\textmd{2D diagram showing a target batch interacting with a source cluster. 
The target batch of radius $r_B$ is a distance R from the source cluster of radius $r_C$.
If $(r_B+r_C)/R<\theta$ and $(n+1)^3 < N_S$, this interaction 
will be approximated using Eq.~\eqref{eqn:particle-cluster-approx-rearranged}.} 
}
\label{fig:batch-cluster-diagram}
\end{figure}

In this work the particle-cluster interactions $u({\bf r}_i,C)$
are organized into batch-cluster interactions.
For a given batch of target particles,
the decision on whether to apply the approximation in Eq.~\eqref{eqn:particle-cluster-approx-rearranged}
is controlled by
the multipole acceptance criterion (MAC) which in this work has the form,
\begin{equation} \label{eqn:MAC}
    \frac{r_B + r_C}{R} < \theta, \quad (n+1)^3 < N_S,
\end{equation}
where $r_B$ is the target batch radius,
$r_C$ is the source cluster radius,
$R$ is the distance between the target batch center and source cluster center,
$n$ is the interpolation degree,
and $N_S$ is the number of source particles in the cluster.
The first part of the MAC ensures the approximation's accuracy and is diagrammed in Fig.~\ref{fig:batch-cluster-diagram},
while the second part ensures its efficiency.

For the interpolation points, the BLTC uses Chebyshev points of the 2nd kind,
\begin{equation}
s_k = \cos\theta_k, \quad \theta_k = \pi k/n, \quad k = 0,\ldots, n,
\end{equation}
which are defined on the interval $[-1,1]$
and
are linearly mapped to clusters located elsewhere.
In addition the BLTC uses the barycentric form of the 
Lagrange polynomials~\cite{berrut-trefethen-2004},
\begin{equation}
\label{eqn:Lagrange_polynomials}
L_k(t) = 
\frac{\displaystyle \frac{b_k}{t-s_k}}{\displaystyle \sum_{k=0}^n \frac{b_k}{t-s_k}}, \quad 
k = 0,\ldots,n,
\end{equation}
where due to the scale-invariance of this form~\cite{salzer-1972},
the barycentric weights are  
\begin{equation}
\label{eqn:barycentric_weights}
b_k = (-1)^k\delta_k, \quad
\delta_k = 
\begin{cases}
1, & k = 1,\ldots,n-1, \cr 
1/2, & k=0,n. \cr
\end{cases}
\end{equation}

\begin{figure}[htb]
\begin{subfigure}[b]{0.4\textwidth}
\centering
\includegraphics[width=\textwidth]{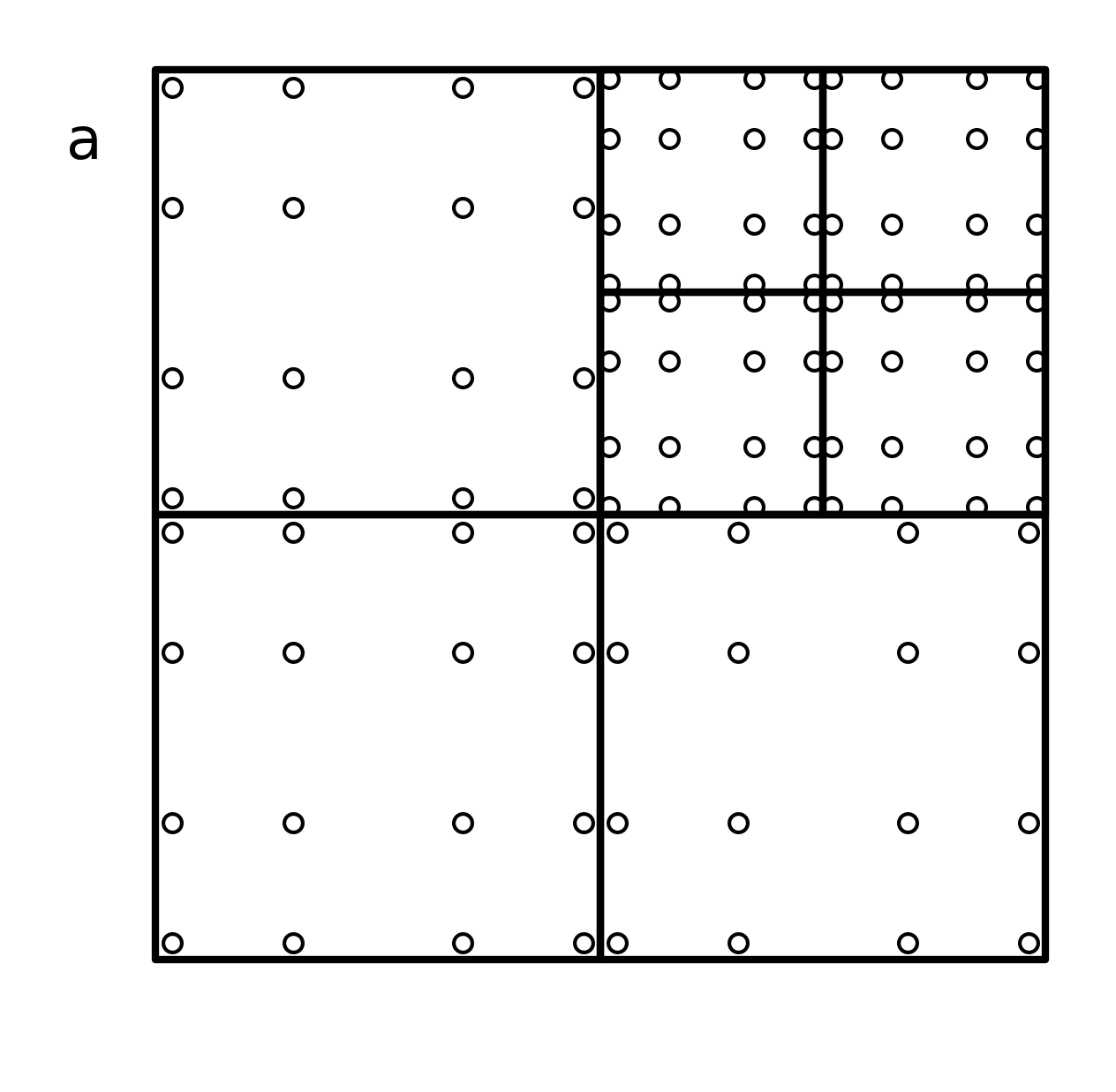}
\end{subfigure}
\raisebox{3.6\height}{\LARGE$\xrightarrow{\text{Eq.}~\eqref{eqn:particle-cluster-approx-rearranged}}$}
\begin{subfigure}[b]{0.4\textwidth}
\centering
\includegraphics[width=\textwidth]{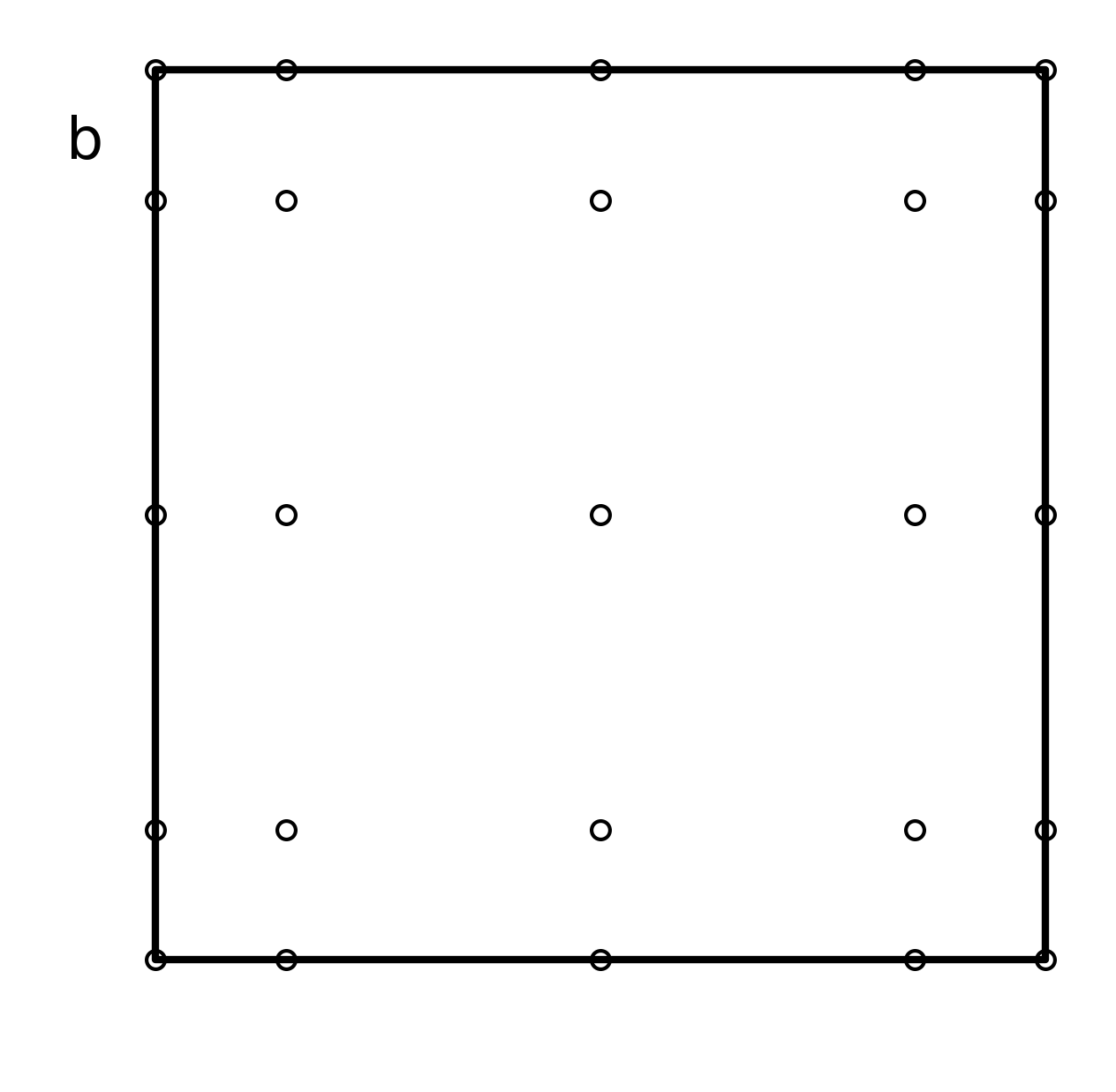}
\end{subfigure}
\caption{
\textmd{ Example of a cluster $C$ in 2D, 
(a) source particles ${\bf r}_j$ in $C$
(these are quadrature points in the adaptive mesh, here defined with order $p=3$),
(b) Chebyshev grid of interpolation points ${\bf s}_{\bf k}$ 
(here defined with degree $n=4$,
these represent the cluster through the particle-cluster approximation 
in Eq.\eqref{eqn:particle-cluster-approx-rearranged}).
}}
\label{fig:treecode-diagram}
\end{figure}

Figure~\ref{fig:treecode-diagram} shows a 2D example of a cluster $C$
comprised of seven quadrature cells;
Fig.~\ref{fig:treecode-diagram}a shows the source particles $\mathbf{r}_j$ in $C$
(these are quadrature points in the adaptive mesh, here defined with order $p=3$),
and
Fig.~\ref{fig:treecode-diagram}b shows the Chebyshev grid of interpolation points 
$\mathbf{s_k}$ in $C$
(here defined with degree $n=4$,
these represent the cluster through the particle-cluster approximation 
in Eq.\eqref{eqn:particle-cluster-approx-rearranged}).
The treecode has two options for computing particle-cluster interactions;
the direct sum in Eq.~\eqref{eqn:direct-sum_2} requires $O(N_S)$ operations,
where $N_S$ is the number of source particles in $C$,
while the approximation in Eq.~\eqref{eqn:particle-cluster-approx-rearranged}
requires $O(n^3)$ operations for interpolation of degree $n$;
hence the approximation is more efficient when $N_S$ is large and $n$ is small,
as in the example in Fig.~\ref{fig:treecode-diagram}.

\subsection{Treecode Algorithm}
\label{section:treecode-algorithm} 

The treecode is described in Algorithm~\ref{alg:treecode}.
The input consists of the quadrature points ${\bf r}_i$, weights $f_i, w_i$,
and the treecode parameters including
MAC $\theta$, 
interpolation degree $n$, maximum leaf size $N_L$, maximum batch size $N_B$.
Line 1 builds the tree of source clusters 
and
the set of target batches.
Line 2 computes the approximation weights $\widehat{f}_{\bf k}$ for each source cluster.
Line 3 computes the batch-cluster interactions for each target batch 
via the recursive function \textsc{ComputePotential};
if the MAC in Eq.~\eqref{eqn:MAC} is satisfied, 
then the approximation is computed with Eq.~\eqref{eqn:particle-cluster-approx-rearranged};
if the MAC fails because $(r_B+r_C)/R \ge \theta$, then there are two options;
if the cluster is a leaf, then the batch interacts directly with the cluster by Eq.~\eqref{eqn:direct-sum_2},
otherwise, the cluster has children 
and the batch interacts with each child;
if the MAC fails because $(n+1)^3 \ge N_S$,
then the batch interacts directly with the cluster by Eq.~\eqref{eqn:direct-sum_2}.
The treecode operation count is $O(N_m\log N_m)$,
where the factor $N_m$ is the number of target particles
and
the factor $\log N_m$ is the number of levels in the tree.

\begin{algorithm}[htb]
\caption{Treecode}\label{alg:treecode}
{\bf input}: quadrature points ${\bf r}_i$ and weights $f_i, w_i, i=1,\ldots,N_m$ \\
{\bf input}: treecode MAC parameter $\theta$, interpolation degree $n$, maximum leaf size $N_L$ \\
{\bf output}: approximate potential $u_i, i=1,\ldots,N_m$
\begin{algorithmic}[1]
\State build tree of source clusters $\{C\}$ and set of target batches $\{B\}$
\State for each source cluster, compute approximation weights $\widehat{f}_{\bf k}$ in Eq.~\eqref{eqn:particle-cluster-approx-rearranged}
\State for each target batch, \textsc{ComputePotential}($B$, root\_{\rm cluster}) \\
\State {\bf function} \textsc{ComputePotential}($Batch$, $Cluster$)
\State \quad if MAC is satisfied, compute batch-cluster approximation by Eq.~\eqref{eqn:particle-cluster-approx-rearranged}
\State \quad else if $(r_B+r_C)/R \ge \theta$ 
\State \quad \quad if $Cluster$ is a leaf, 
compute batch-cluster interaction by direct sum in Eq.~\eqref{eqn:direct-sum_2}
\State \quad \quad else for each $Child$ of $Cluster$, 
\textsc{ComputePotential}($Batch$, $Child$)
\State \quad else if $(n+1)^3 \ge N_S$
\State \quad \quad
compute batch-cluster interaction by direct sum in Eq.~\eqref{eqn:direct-sum_2}
\State {\bf end function}
\end{algorithmic}
\end{algorithm}


\subsection{Treecode Accuracy} \label{section:Treecode-Approximation-Error}

\begin{figure}[htb]
\centering
\includegraphics[width=0.5\textwidth]{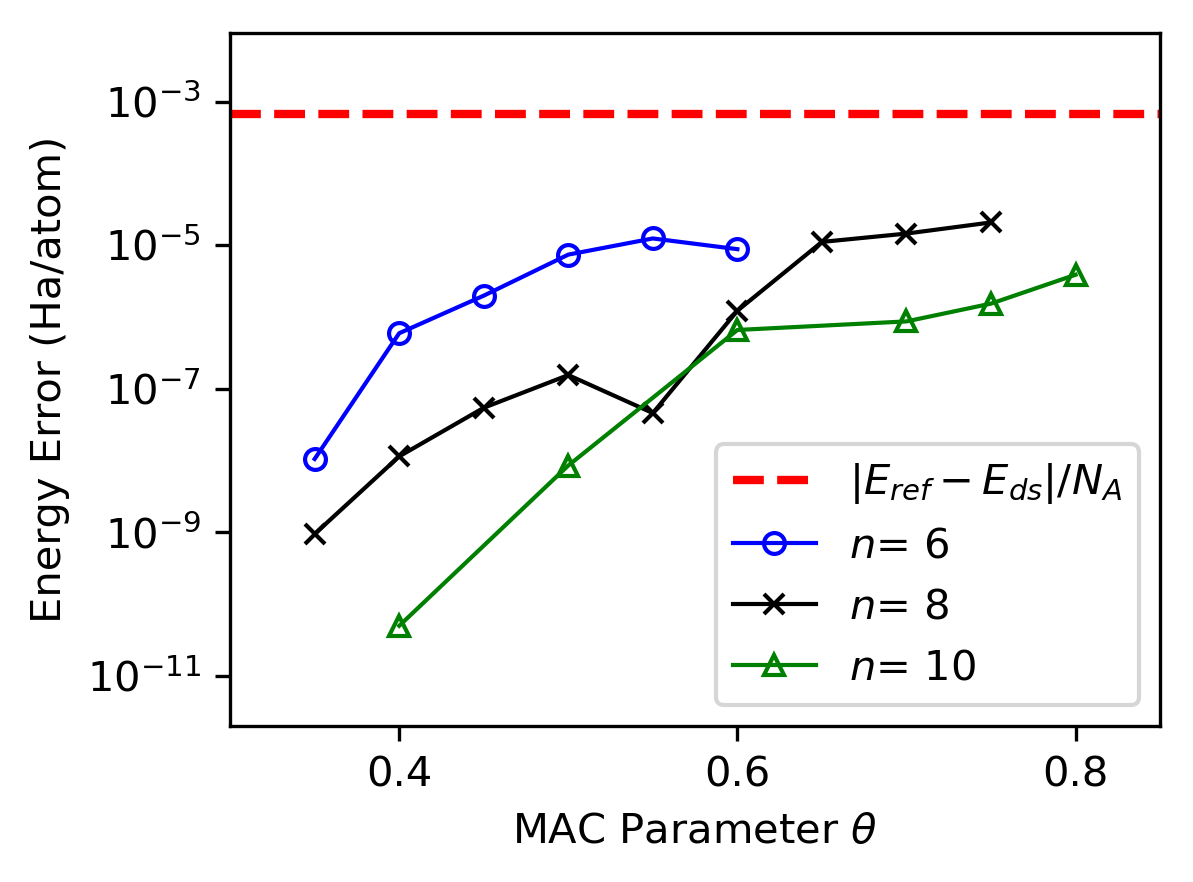}
\caption{
\textmd{Comparison of the treecode approximation error to the underlying discretization error for the Carbon monoxide molecule.
The mesh contains $N_m = 661625$ points and results in a discretization error of $|E_{ref} - E_{ds}|/N_A =$ 6.56e$-4$ Ha/atom (red dashed line).
Solid curves and symbols show the treecode approximation error $|E_{ds} - E_{tc}|/N_A$
for treecode MAC parameter $0.35 \le \theta \le 0.8$ and interpolation degree $n=6,8,10$.
}
}
\label{fig:treecode-approx-error}
\end{figure}

The sum in Eq.~\eqref{eqn:direct-sum} is a discretization
of the convolution integral in Eq.~\eqref{eqn:convolution_integral},
and
it is important to ensure that the treecode approximation error
is less than discretization error.
We document the accuracy of the treecode for the carbon monoxide molecule 
with domain $[-20,20]^3$ a.u., 
quadrature order $p=4$, 
mesh refinement tolerance $tol_m=$3e$-7$, 
SCF tolerance $tol_{scf}=$1e$-5$,
and
Green Iteration tolerance $tol_{gi}=$1e$-6$.
In this case the number of mesh points is $N_m = 661625$.
We compute the ground-state energy with and without the treecode,
and
record the discretization error $|E_{ref} - E_{ds}|/N_A$,
and
treecode approximation error $|E_{ds} - E_{tc}|/N_A$,
where 
$E_{ref}$ is the reference energy computed by DFT-FE,
and
$E_{ds}, E_{tc}$ are computed by the present method using direct summation
and
the treecode, respectively.
Throughout this work the source clusters and target batch size parameters are set to $N_L=N_B=2000$.
Figure~\ref{fig:treecode-approx-error} shows that the
discretization error of the present method is $|E_{ref} - E_{ds}|/N_A =$ 6.56e$-4$ Ha/atom (red dashed line),
while the treecode approximation error is much smaller.
This confirms that in this range of parameter values,
the numerical errors introduced by the treecode do not upset the
accuracy of the discretization.

\subsection{Treecode Efficiency on a 6-core CPU and single GPU} 
\label{section:GPU-implementation}

This section documents the BLTC efficiency 
on a 6-core 2.6 GHz Intel Core i7 processor
and
a single NVIDIA Titan V GPU.
The treecode was programmed in C,
with OpenMP directives for parallelizing over multiple CPU cores 
and OpenACC directives for running on the GPU.
As noted above,
the GPU implementation takes advantage of the fact that the 
particle-cluster interaction in Eq.~\eqref{eqn:direct-sum_2}
and
the approximation in Eq.~\eqref{eqn:particle-cluster-approx-rearranged}
both have the same direct sum structure involving independent kernel evaluations.
The GPU processes the particle-particle interactions 
between the target batch and source cluster in parallel without thread divergence;
this is because the MAC applies uniformly to all particles in a 
given target batch~\cite{vaughn-wilson-krasny-2020}.
In practice, interaction lists are precomputed for each target batch 
to identify the source clusters interacting with the batch.

The performance of the BLTC on both platforms is demonstrated
by computing the Hartree energy $E_H$ in Eq.~\eqref{eqn:electrostatics}
for the carbon monoxide molecule
using the electron density from the first SCF iteration.
Results are shown in Table~\ref{table:treecode-and-gpu-speedup} 
using direct summation and the treecode,
for quadrature order $p=4$
and 
mesh refinement parameter $tol_m$ between 1e$-3$ and 1e$-8$
yielding the indicated mesh size $N_m$.
The direct sum energy values in the 3rd column converge as the mesh is refined, 
and
the 4th column shows the corresponding discretization error
using the value $E_{H} = 74.88578$ obtained with $tol_m$=1e-8 as the reference.
The 5th column records the treecode approximation error
for MAC $\theta=0.7$ and degree $n=8$;
this is the difference between the value of $E_H$
computed by direct summation (column 3) 
and 
the value computed by the treecode (not shown).
The results show that the treecode approximation error is well below the 
discretization error and within chemical accuracy.

The remainder of Table~\ref{table:treecode-and-gpu-speedup}
records computation times on the 6-core CPU and GPU
for direct summation (ds) and the treecode (tc).
Averaging over the six runs in Table~\ref{table:treecode-and-gpu-speedup},
direct summation runs 192 times faster on the GPU than on the 6-core CPU,
while the treecode runs 70 times faster.
On both platforms
the treecode is faster than direct summation,
and
the speedup (ds/tc) increases as the mesh is refined; 
this is consistent with
$O(N_m^2)$ scaling for direct summation
and
$O(N_m\log N_m)$ scaling for the treecode.
In particular,
for the largest mesh size with approximately 2.2 million mesh points,
the treecode computation time on the GPU is less than 18~s,
which is about 4.5 times faster than direct summation.

\begin{table}[htb]
\centering
\small
\begin{tabular}{ccccc|ccc|ccc}
\textbf{} & \textbf{} & \textbf{} & \textbf{} & \textbf{} &
\multicolumn{3}{c|} {6-core CPU time (s)} & 
\multicolumn{3}{c} {GPU time (s)} \\ \cline{6-11}
$tol_m$ & $N_m$ & $E_H$ (Ha) & ds error & tc error & ds & tc & ds/tc & ds & tc & ds/tc \\ \hline
1e-3 & 141000 & 74.88640 & 6.20e-4 & 5.31e-7 & 70.42 & 26.24 & 2.68 & 0.39 & 0.47 & 0.82 \\
1e-4 & 184750 & 74.87690 & 8.88e-3 & 1.20e-6 & 150.13 & 41.40 & 3.63 & 0.67 & 0.68 & 0.97 \\
1e-5 & 249500 & 74.88668 & 9.00e-4 & 8.85e-6 & 216.17 & 80.32 & 2.69 & 1.13 & 1.13 & 1.00 \\
1e-6 & 459500 & 74.88551 & 2.70e-4 & 3.68e-5 & 638.23 & 196.11 & 3.25 & 3.57 & 2.72 & 1.31 \\
1e-7 & 928500 & 74.88574 & 4.00e-5 & 6.33e-6 & 2509.2 & 486.54 & 5.16 & 13.70 & 6.49 & 2.11 \\
1e-8 & 2224375 & 74.88578 & na & 2.57e-6 & 15239.6 & 1373.90 & 11.09 & 78.31 & 17.27 & 4.54
\end{tabular}
\caption{
\textmd{Treecode accuracy and acceleration for the Carbon monoxide molecule using quadrature order $p=4$ and mesh refinement parameter $tol_m$,
giving mesh size $N_m$, 
Hartree energy $E_H$ (Ha) in Eq.~\eqref{eqn:electrostatics} for
electron density in first SCF iteration,
ds error (discretization error, computed using $tol_m$ = 1e-8 as reference),
tc error (treecode error, $|E_H({\rm ds}) - E_H({\rm tc})|$ using MAC $\theta=0.7$ and interpolation degree $n=8$).
Run time~(s) for the direct sum (ds) and treecode (tc) and treecode speedup (ds/tc) on a 6-core CPU and a single GPU}
}
\label{table:treecode-and-gpu-speedup}
\end{table}


\subsection{BLTC Parallel Efficiency on a Single GPU node with 1, 2 or 4 GPUs}

\begin{figure}[htb]
\centering
\begin{subfigure}[b]{0.325\textwidth}    
\centering 
\includegraphics[width=\textwidth]{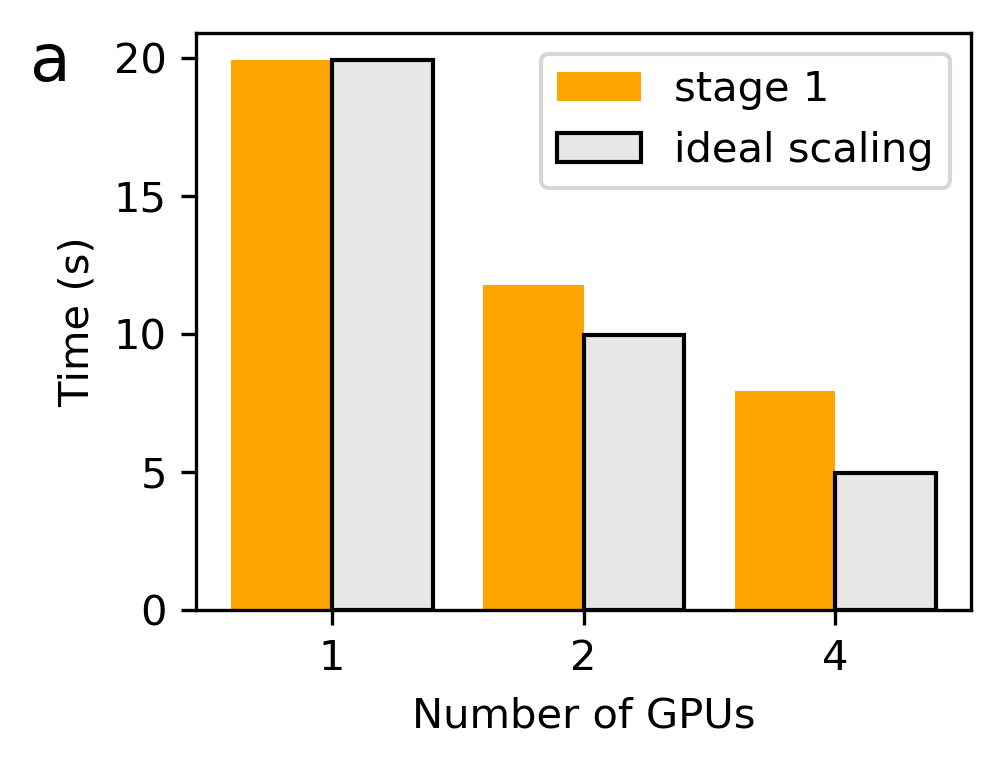}
\end{subfigure}
\begin{subfigure}[b]{0.325\textwidth}    
\centering 
\includegraphics[width=\textwidth]{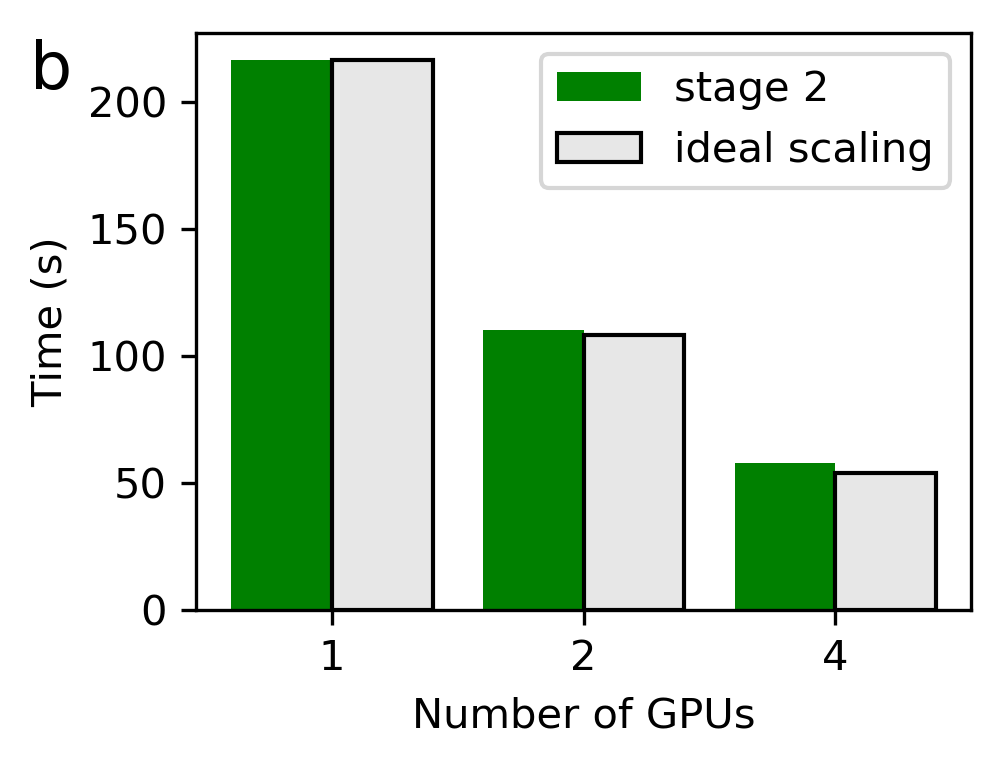}
\end{subfigure}
\begin{subfigure}[b]{0.325\textwidth}    
\centering 
\includegraphics[width=\textwidth]{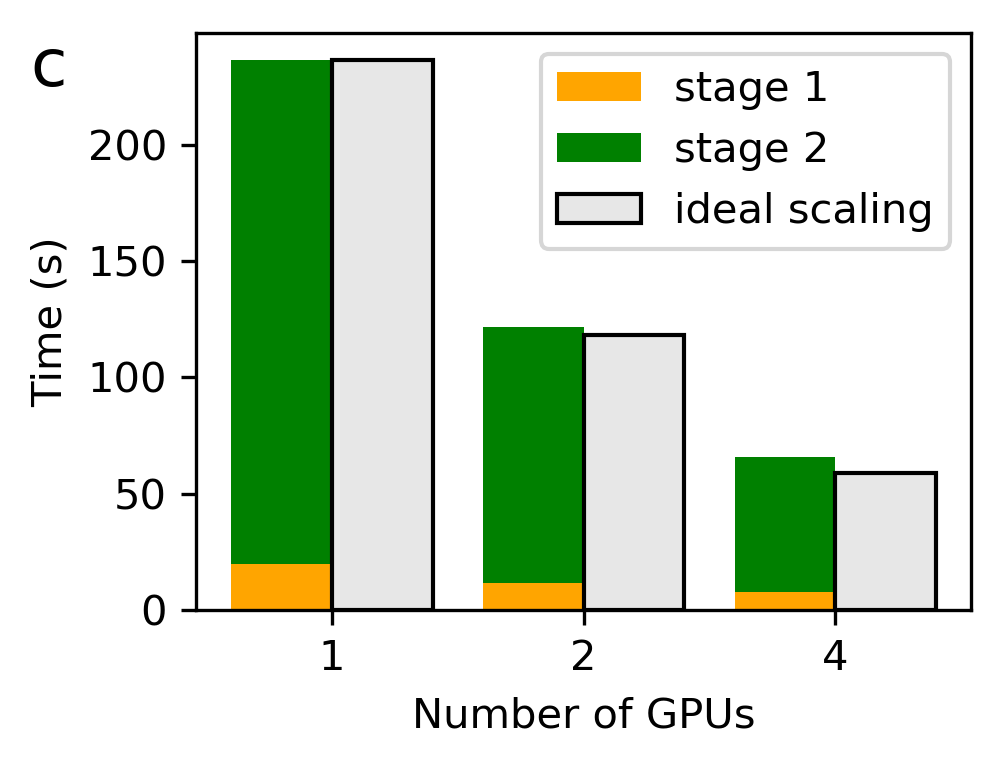}
\end{subfigure}
\caption{
\textmd{Parallel efficiency of BLTC on a single GPU node for
10 million particles interacting via Coulomb kernel. 
Treecode MAC parameter $\theta=0.7$ and interpolation degree $n=7$ 
yield treecode approximation error 2.31e-06 
($L_2$ error with respect to direct sum).
The results show computation time~(s) 
and ideal scaling time~(s) using 1, 2 and 4 GPUs for 
(a) stage 1 (precompute),
(b) stage 2 (compute),
and (c) total time.
For comparison, the direct sum time on 4~GPUs is 1668~s.}
}
\label{fig:parallel-scaling}
\end{figure}

This subsection documents the parallel efficiency
of the BLTC on a single GPU node running with 1, 2 or 4 GPUs,
using OpenMP to parallelize across GPUs
with one thread assigned to each GPU.
The test system has 10 million particles randomly located in a cube
interacting via the Coulomb kernel.  
The work is divided into two stages;
stage~1 encompasses the precomputing tasks in lines 1-2 of Algorithm~\ref{alg:treecode}
and
stage~2 encompasses the batch-cluster computing in line 3.
Figure~\ref{fig:parallel-scaling} 
shows the parallel efficiency of each stage and the entire computation 
as the number of GPUs increases from 1 to 4.
The precompute stage scales less efficiently than the compute phase, 
due to some serial computation embedded in these tasks
(85\% on 2~GPUs, 63\% on 4~GPUs),
but this accounts for only a small fraction of the total computation time.
The compute stage has close to ideal scaling
(98\% on 2~GPUs, 94\% on 4~ GPUs);
moreover this stage accounts for a large fraction of the total computation time
and
therefore the treecode achieves 90\% efficiency for the entire computation on 4~GPUs.  

\section{Ground State Energy Computations for Atoms and Molecules} 
\label{section:Total-Energy-Accuracy}

The ground-state energy of several atoms (Li, Be, O) 
and 
small molecules (${\rm H}_2$, CO, ${\rm C}_6{\rm H}_6$)
was computed using treecode-accelerated Green Iteration (TAGI) with
the LDA exchange-correlation functional~\cite{ceperley-alder-1980,perdew-zunger-1981}.
For each system the SCF iteration continued until the 
density residual fell below $tol_{scf}=$1e$-4$.
$tol_{gi}$ was set to 3e$-3$ for the first step in the SCF, 
then gradually reduced to 1e$-5$ over the next four steps.
The TAGI discretization parameters 
(quadrature order $p$, adaptive mesh parameter $tol_m$)
and
treecode parameters (degree $n$, MAC $\theta$)
were chosen to ensure chemical accuracy of 1~mHa/atom in the computed ground-state energy.
The computations were performed on a single node,
where the treecode (written in C with OpenMP+OpenACC) was run on the four GPUs
and the remainder of the code (written in Python) was run in serial on one CPU core.

Table~\ref{table:timings} presents the parameters and results for each system.
The numerical parameters ($p, tol_m, n, \theta$) are chosen
to ensure chemical accuracy in the energy,
and
the heavier carbon and oxygen atoms 
require somewhat higher numerical resolution
than the lighter hydrogen, lithium, and beryllium atoms.
In particular,
a larger mesh size $N_m$ requires slightly tighter treecode parameters 
(increasing degree from $n=6$ to $n=7,8$, 
decreasing MAC from $\theta = 0.8$ to $\theta = 0.7,0.6$). 
Column 9 records the error in the ground-state energy computed by TAGI with respect to reference energies computed to $0.1$~mHa/atom accuracy with DFT-FE~\cite{motamarri-das-rudraraju-ghosh-davydov-gavini-2019},
showing that TAGI achieves chemical accuracy.
Column 10 records the total wall clock computation time~(s).
The benzene molecule (C$_6$H$_6$, $N_e=42$) is the largest system considered;
the computation used approximately 1.5 million mesh points
and
required less than 4 hours of wall clock time.

\begin{table}[htb]
\centering
\small
\begin{tabular}{cccccccccc}
system & $N_e$& $p$ & $tol_m$ & $N_m$ & $n$ & $\theta$ & $E_{TAGI}$ & error (Ha/atom) & time (s) \\ 
\hline
Li  & 3 & 4 & 7e-6 & 232000 & 6 & 0.8 & -7.334051 & 4.59e-4 & 35.75 \\
Be  & 4 & 3 & 1e-5 & 179712 & 6 & 0.8 & -14.445658 & 5.32e-4 & 29.35 \\
O   & 8 & 4 & 3e-7 & 421000 & 6 & 0.8 & -74.469668 & -3.38e-4 & 143.09 \\
& &  &  &  &  &  &  & \\
H$_2$ & 2 & 3 & 1e-3 & 51200 & 6 & 0.8 & -1.13584 & 9.05e-4 & 6.67 \\
CO & 14 & 4 & 3e-7 & 661625 & 7 & 0.7 & -112.473372 & -7.21e-4 & 932.03 \\
C$_6$H$_6$ & 42 & 4 & 3e-6 & 1464500 & 8 & 0.6 & -230.193158 & -3.64e-4 & 13947.89 
\end{tabular}
\caption{
\textmd{Ground-state energy computations of atoms and small molecules using TAGI with errors computed with respect to reference values $E_{ref}$ computed using 
DFT-FE~\cite{motamarri-das-rudraraju-ghosh-davydov-gavini-2019}.
TAGI discretization parameters 
(quadrature order $p$, adaptive mesh parameter $tol_m$, mesh size $N_m$),
treecode parameters (degree $n$, MAC $\theta$);
error (Ha/atom) and total wall clock computation time~(s)
on a single node with 4 GPUs.} 
}
\label{table:timings}
\end{table}

\begin{figure}[htb]
\centering
\begin{subfigure}[b]{0.45\textwidth}
\centering
\includegraphics[width=0.65\textwidth]{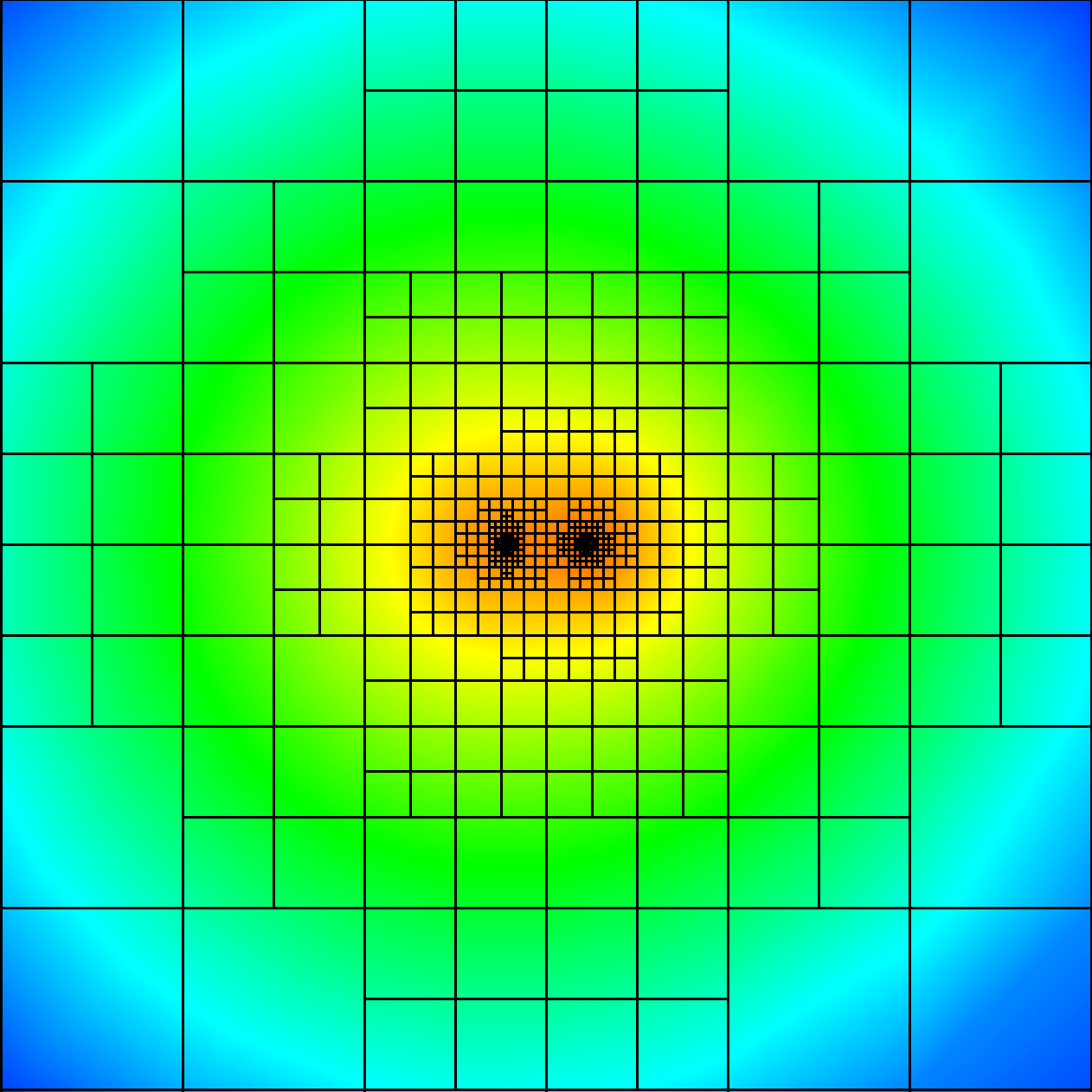}
\includegraphics[width=0.3\textwidth]{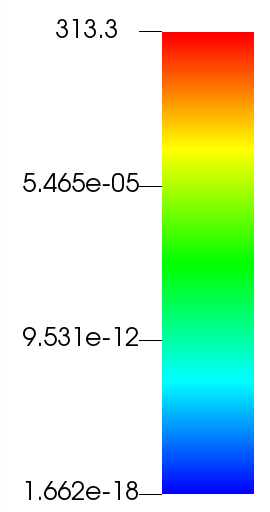}
\end{subfigure}
\hskip 0.2in
\begin{subfigure}[b]{0.45\textwidth}
\centering
\includegraphics[width=0.65\textwidth]{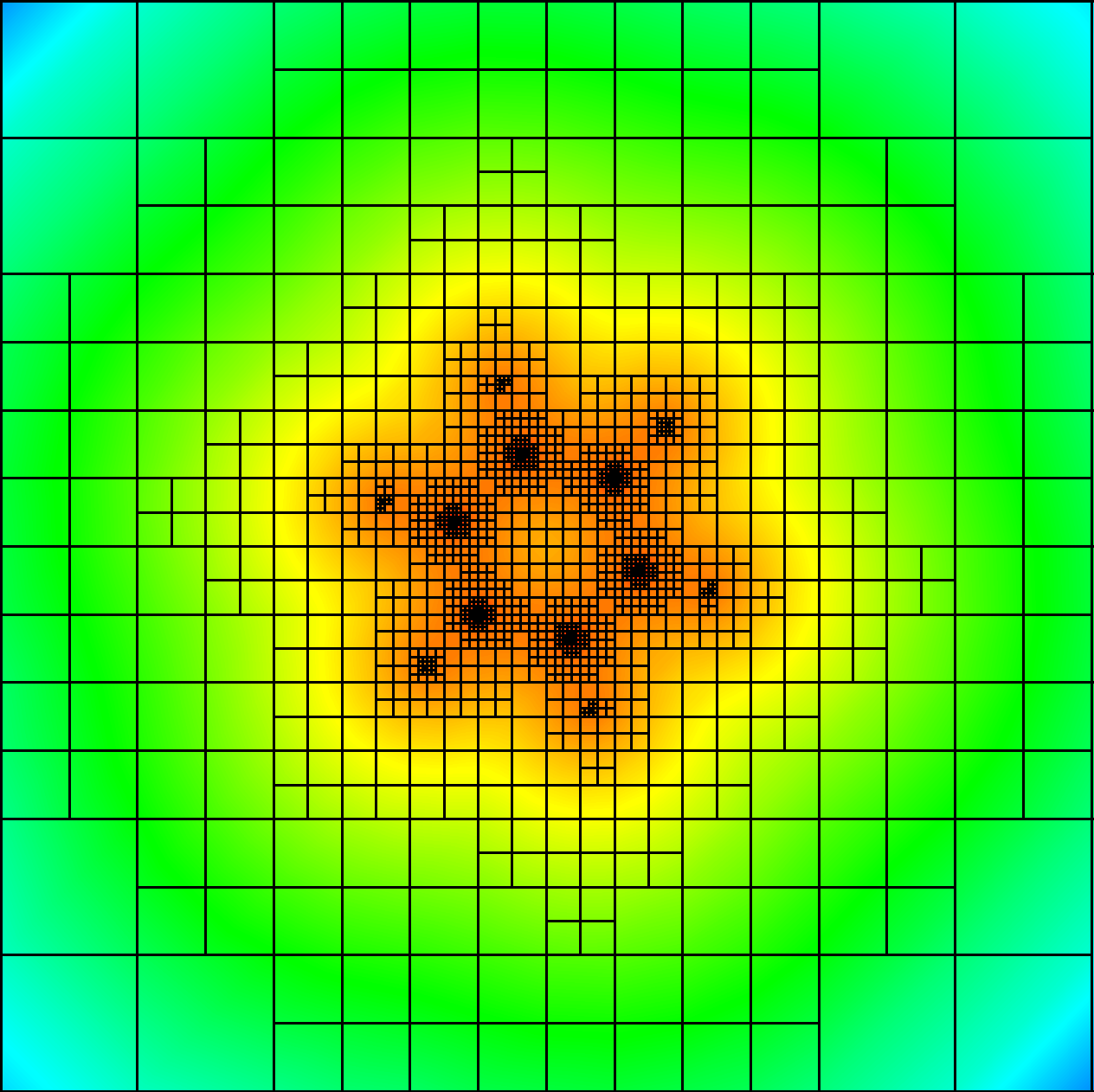}
\includegraphics[width=0.3\textwidth]{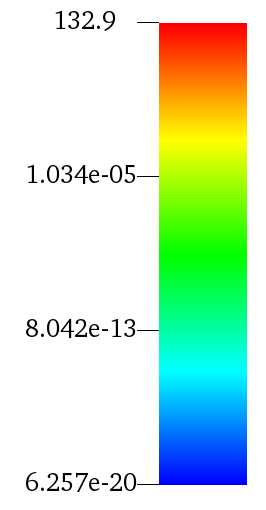}
\end{subfigure}
\caption{
\textmd{2D slices of the adaptively refined mesh 
and 
computed electron density $\rho$ for 
(a) carbon monoxide molecule (CO), 
and (b) benzene molecule (C$_6$H$_6$).
The slices are taken at $z=0$ and show $[-15,15]^2$~a.u. in $xy$-plane.}
}
\vskip -175pt
\setlength{\unitlength}{1cm}
\begin{picture}(5,1)
\put( -5.45, 0.50){\large{\bf a}}
\put(  2.50, 0.45){\large{\bf b}}
\end{picture}
\vskip 140pt
\label{fig:benzene-density}
\end{figure}

Figure~\ref{fig:benzene-density} show 2D slices of the 
adaptively refined mesh and computed electron density 
for the carbon monoxide molecule~\ref{fig:benzene-density}a 
and benzene molecule~\ref{fig:benzene-density}b. 
Both molecules are located in the z=0 plane; 
for the CO molecule, the carbon atom is at (-1.06581,0,0) 
and the oxygen atom is at (+1.06581,0,0);
for the benzene molecule, the six carbon atoms are at 
$(\mp0.682781, \pm2.548170, 0)$, 
$(\pm2.548230, \mp0.682767, 0)$,
$(\pm1.86544, \pm1.86541,  0)$, 
and the six hydrogen atoms are at
$(\mp1.21247, \pm4.52502,  0)$,
$(\pm4.52548, \mp1.21333, 0)$,
$(\pm3.31252, \pm3.31351,  0)$.
The slices are taken in the $z=0$ plane
and show the region $[-15,15]$ a.u. in the $xy$-plane.
The adaptive mesh successfully captures the variation
in magnitude of the electron density.

\section{Conclusions}
\label{section:Conclusion}

We presented a real-space method for all-electron Kohn-Sham DFT computations 
called Treecode-Accelerated Green Iteration (TAGI).
TAGI is based on a reformulation of the Kohn-Sham equations
in which the eigenvalue problem 
for the energies and wavefunctions $(\varepsilon_i,\psi_i)$
in differential form 
is recast as a fixed-point problem in integral form
by convolution with the bound state Helmholtz 
Green's function~\cite{kalos-1962}.
In each SCF iteration the fixed-points are computed by Green Iteration, 
where the convolution integrals are discretized on an adaptive mesh
and
the discrete convolution sums are efficiently evaluated using a 
GPU-accelerated treecode. 

TAGI relies on several key techniques to achieve chemical accuracy 
and computational efficiency.
First, 
the Fej\'er quadrature rule
and
adaptive mesh refinement based on integration of a test function
are used to compute integrals and represent the fields. 
Second, singularity subtraction is applied to 
evaluate convolution integrals having singular kernels;
in particular a standard scheme is used for the Yukawa kernel 
in the Green Iteration convolutions,
and
we  developed a new scheme for the Coulomb kernel
in the Hartree potential convolutions
since the standard scheme is inapplicable in that case.
Third, a gradient-free method is employed to update the eigenvalues in Green Iteration.
Fourth, the fixed-point iteration for the wavefunctions and eigenvalues in
Green Iteration is accelerated using Anderson mixing.
Fifth, the discrete convolution sums are computed efficiently
using a barycentric Lagrange treecode (BLTC),
which reduces the operation count from $O(N_m^2)$ to $O(N_m\log N_m)$,
where $N_m$ is the number of mesh points.
The GPU implementation of the BLTC is facilitated by the
properties of barycentric Lagrange interpolation
including its scale-invariance 
and
the fact that the particle-cluster approximation
in Eq.~\eqref{eqn:particle-cluster-approx-rearranged}
consists of independent kernel evaluations
which can be evaluated concurrently~\cite{vaughn-wilson-krasny-2020}.

We demonstrated the effect of these techniques on the carbon monoxide molecule.
First, we investigated the quadrature rule
and
adaptive mesh refinement scheme,
showing that the ground-state energy of the CO molecule is computed to
within chemical accuracy of 1 mHa/atom 
using roughly 600,000 quadrature points
and 4th order quadrature.
Second, we demonstrated the effect of the singularity subtraction schemes;
the ground-state computation was performed with and without singularity subtraction 
on a sequence of progressively refined meshes, 
and 
we observed a 100-fold reduction in error using singularity subtraction.
Third, we compared three methods for updating the eigenvalues in Green Iteration, 
1) Laplacian update using the Rayleigh quotient of the Hamiltonian 
differential operator, 
2) gradient update using integration by parts to reduce the order of the operator, 
and 
3) gradient-free update;
on a wide range of meshes the gradient-free update yields
a 10-fold improvement in accuracy over the gradient update, 
and 
a 100-fold improvement over the Laplacian update.
Fourth, 
we investigated the convergence of Green Iteration in the first SCF iteration
for the CO molecule,
showing that the spectral gap in the Hamiltonian eigenvalues controls the 
rate of convergence of the eigenfunctions in Green Iteration;
in particular,
a small gap $|\varepsilon_i - \varepsilon_{i+1}|$ implies 
slow convergence of the eigenfunction $\psi_i$;
the convergence rates were predicted
and
then verified computationally;
finally we showed in the case of slow convergence,
Green Iteration can be accelerated by applying Anderson mixing to the eigenpairs,  
yielding a 6-fold reduction in the number of iterations in this example.
Fifth, 
we demonstrated the treecode's ability to 
rapidly compute accurate approximations of the discrete convolution sums;
in ground-state computations for the CO molecule, 
we showed that the treecode approximation error can be driven below the 
discretization error;
we then demonstrated the speedup of the treecode over 
direct summation on both a 6-core CPU (parallelized with OpenMP) 
and a GPU (parallelized with OpenACC),
achieving an 11$\times$ speedup on the CPU 
and a 4.5$\times$ speedup on the GPU;
finally we observed a 70$\times$ speedup of the treecode running on the GPU
in comparison with the CPU.

We then performed TAGI computations for several atoms 
and 
small molecules
on a single node with 4~GPUs,
and 
verified the accuracy of the ground-state energy with respect
to reference values. 
The results demonstrate the chemical accuracy and computational efficiency 
afforded by TAGI on these benchmark systems.

In future work we will implement a distributed memory version of TAGI
able to run on multiple nodes,
relying on domain decomposition to treat larger systems.
We will also consider a treecode based on barycentric Hermite interpolation, 
to more efficiently reach high accuracy regime needed in this application~\cite{Krasny:2019aa}.
In addition, 
we will extend TAGI to pseudopotential computations using 
Optimized Norm-Conserving Vanderbilt (ONCV) pseudopotentials~\cite{hamann-2013}; 
this will alleviate the need for significant mesh refinement near atomic nuclei 
and 
reduce the number of wavefunctions per atom, 
allowing TAGI to scale to larger systems.


\section*{Acknowledgement}
This work was supported by National Science Foundation grant DMS-1819094, 
and 
the 
Michigan Institute for Computational Discovery and Engineering (MICDE) 
and
Mcubed program at the University of Michigan.
RK thanks Lunmei Huang and Li Wang for early discussions on the Green's function approach to DFT.
NV thanks Bikash Kanungo and Sambit Das for providing reference values and the single-atom radial data.
Computational resources and services were provided by Advanced Research Computing-Technology Services (ARC-TS) at the University of Michigan.

\bibliography{library}
\appendix

\section{Symbols}
\label{appendix:parameters}

\begin{table}[htb]
\begin{tabular}{ll}
Variables, fields, and operators  & Symbol \\ \hline
Electron density, Eq.~\eqref{eqn:density-construction}       & $\rho$ \\
Effective potential, Eq.~\eqref{eqn:effective-potential} & $V_{eff}[\rho]$ \\
Hamiltonian operator, Eq.~\eqref{eqn:Kohn-Sham-differential} & $\mathcal{H}[\rho]$ \\
Hamiltonian eigenpairs, Eq.~\eqref{eqn:Kohn-Sham-differential}    & $(\varepsilon_i,\psi_i) $  \\
Green Function Integral operator, Eq.~\eqref{eqn:integral_operator}     & $\mathcal{G}(\varepsilon)$ \\
Integral eigenpairs, Eq.~\eqref{eqn:mu-phi-equation} & $(\mu_i,\phi_i)$ \\
\\
Physical parameters and constants  & Symbol \\ \hline
Number of atoms, Eq.~\eqref{eqn:classical-electrostatics}    & $N_A$             \\
Atomic positions, Eq.~\eqref{eqn:classical-electrostatics} & $\mathbf{R}_j$ \\
Nuclear charges, Eq.~\eqref{eqn:classical-electrostatics} & $Z_j$ \\
Number of electrons, Eq.~\eqref{eqn:Fermi}    & $N_e$ \\
Boltzmann constant, Eq.~\eqref{eqn:density-construction}  &  $k_B$ \\
Temperature, Eq.~\eqref{eqn:density-construction}  & $T$\\
Fermi Energy, Eq.~\eqref{eqn:Fermi}     & $\mu_F$\\
\\
Numerical Parameters          & Symbol   \\ \hline
Number of wavefunctions, Eq.~\eqref{eqn:Fermi}     & $N_w$             \\
Mesh refinement parameter, Eq.~\eqref{eqn:refinement_criterion}    & $tol_m$\\
Number of cells, Eq.~\eqref{eqn:quadrature-example}    & $N_c$             \\
Quadrature order, Eq.~\eqref{eqn:quadrature-example}  & $p$               \\
Number of quadrature points, Eq.~\eqref{eqn:quadrature-example} & $N_m$             \\
SCF convergence tolerance, Alg.~\eqref{alg:SCF}   & $tol_{scf}$       \\
Green Iteration convergence tolerance, Alg.~\eqref{alg:GreensIteration}   & $tol_{gi}$ \\
Anderson mixing parameter                           & $\beta$           \\
Gaussian singularity subtraction parameter, Eq.~\eqref{eqn:coulomb-conv-ss} & $\alpha$  \\
Gauge shift, Eq.~\eqref{eqn:gauge-shift}      & $V_{shift}$       \\
Treecode MAC parameter, Eq.~\eqref{eqn:MAC}  & $\theta$          \\
Treecode interpolation degree, Eq.~\eqref{eqn:particle-cluster-approx-rearranged}    & $n$   \\
Treecode maximum batch size                         & $N_B$             \\
Treecode maximum leaf cluster size                  & $N_L$   \\
Number if source particles in cluster                  & $N_S$  
\end{tabular}
\label{table:parameters-table}
\end{table}

\end{document}